\documentclass[aps,showpacs]{revtex4}
\usepackage{graphicx}
\begin{document}
\title{ Gauge invariant non-linear electrodynamics motivated by a spontaneous breaking
of the Lorentz symmetry }
\author{Jorge Alfaro$^{1}$ and Luis F. Urrutia$^{2}$}
\affiliation{$^{1}$ Facultad de F\'\i sica, Pontificia Universidad
Cat\'olica de Chile, Casilla 306, Santiago 22, Chile }
\affiliation{$^{2}$ Instituto de Ciencias Nucleares, Universidad
Nacional Aut{\'o}noma de M{\'e}xico, A. Postal 70-543, 04510
M{\'e}xico D.F.}
\begin{abstract}
We introduce a new version of non-linear electrodynamics which is produced
by a spontaneous symmetry breaking of Lorentz invariance induced by the
non-zero expectation value of the gauge invariant electromagnetic field
strength. The symmetry breaking potential is argued to effectively arise
from the integration of massive gauge bosons and fermions in an underlying
fundamental theory. All possible choices of the vacuum lead only to the remaining
invariant subgroups T(2) and HOM(2). We explore in detail the plane wave
solutions of the linearized sector of the model for an arbitrary vacuum.
They present  two types of dispersion relations. One corresponds to the case of the usual
Maxwell electrodynamics with the standard polarization properties of the fields.  
The other dispersion relation involves
anisotropies determined by the structure of the vacuum. The corresponding
fields reflect these anisotropies. The model is stable
in the small Lorentz invariance violation approximation.
We have also embedded our model in the
photon sector of the Standard Model Extension, in order to translate the
many bounds obtained in the latter into corresponding limits for our
parameters. The one-way anisotropic speed of light is calculated for a
general vacuum and its isotropic component is strongly bounded by $\tilde
\delta c/c < 2 \times 10^{-32}$. The anisotropic violation contribution is
estimated by introducing an alternative definition for the difference of the
two-way speed of light in perpendicular directions, $\Delta c$, that is
relevant to Michelson-Morley type of experiments and which turns out to be
also strongly bounded by $\Delta c/c < 10^{-32}$. Finally, we speculate on
the relation of the vacuum energy of the model with the cosmological
constant and propose a connection between the vacuum fields and the
intergalactic magnetic fields.
\end{abstract}
\maketitle

\section{Introduction}

The possible violation of Lorentz invariance has recently received a
lot of attention, both from the experimental and theoretical sides.
In the latter case, mainly in connection with possible effects
arising from drastic modifications of space-time at distances of the
order of Planck length suggested by most of the current quantum
gravity approaches. Experiments and astrophysical observations set
stringent bounds upon the parameters describing such violations,
which nevertheless are still been subject to improvement in their
precision. Broadly speaking, there are two basic possibilities which
produce such a breaking: (1) one introduces by hand a number of
non-dynamical tensor fields, whose fixed directions induce the
corresponding breaking in a given reference frame. Examples of this
approach are the phenomenological Myers-Pospelov model {\cite{MP}}
together with QED in a constant axial vector background
\cite{aacgs}. (2) a second possibility is to dwell upon spontaneous
symmetry breaking (SSB). This is basically what is done in the
Standard Model Extension \cite{KOSTELECKY0}, \cite{REV}, where such
non-dynamical tensor fields are assumed to arise from vacuum
expectation values of some basic fields belonging to a more
fundamental model, like string theory for example.

Nevertheless, we are interested in studying a model of
electrodynamics which incorporates spontaneous Lorentz symmetry
breaking, without having to go to a string theory setting. In order
to produce such a model we require the presence of the non-zero
vacuum expectation value of a tensorial field of rank greater than
zero.

A first possibility arises from having a non-zero VEV of the
potential field $A_{\mu }$. This case has been thoroughly studied
and leads to the interesting idea that the photon arises as the
corresponding Goldstone boson of the global spontaneous Lorentz
symmetry breaking. Normally the masslessness of the photon is
explained by invoking abelian gauge invariance under the group
$U(1)$. This almost sacred principle has undoubtedly been
fundamental in the development of physics and it is very interesting
to
explore the possibility that it could have a dynamical origin \cite%
{PRLNielsen}. This idea goes back to the works of Nambu \cite{Nambu}
and
Bjorken \cite{Bjorken} together with many other contributions \cite%
{KT,others}. Recently it has been revived in references \cite%
{kostelecky1,chkareulli, azatov,mohapatra,chkareulli1}. One of the
most explored approaches along these lines starts form a theory with
a vector field $B^{\mu }$ endowed with the standard electrodynamics
kinetic term plus a potential designed to break Lorentz symmetry via
a non-zero vacuum expectation value $<B^{\mu }>$, which defines a
preferred direction in space-time. This potential also breaks gauge
invariance. These are the so called bumblebee models
\cite{kostelecky2}. The subsequent symmetry breaking, obtained from
the non-zero minimum of the potential, splits the original four
degrees of freedom in $B^{\mu }$ into three vectorial
Nambu-Goldstone bosons $A^{\mu }$, satisfying the constraint $A_{\mu
}A^{\mu }=\pm M^{2}$, to be identified with the photon, plus a
massive scalar field $\sigma $, which is assumed to be excited at
very high energies. Under this threshold one basically recovers the
Lagrangian for electrodynamics in vacuum plus the above mentioned
non-linear constraint, which is interpreted as a gauge fixing
condition. Calculations at the tree level \cite{Nambu} and to the
one loop level \cite{azatov} have been carried on, showing that the
possible Lorentz violating effects are not present in the physical
observables and that the results are completely consistent with the
standard gauge invariant electrodynamics. This is certainly a
surprising result, which can be really appreciated from the
complexity of the calculation in Ref. \cite{azatov}. Nevertheless,
alternative kinetic terms in the bumblebee models can lead to a
theory differing substantially from electromagnetism
\cite{BLUHMETAL}. The use of nonpolynomial interactions
within this framework has been explored in \cite{altschul2}. The idea of a photon as a Goldstone boson
has also been extended to gravitons in general relativity \cite{kostelecky2,carroll2,potting}.

It is noteworthy to recall here that the
constraint $A_{\mu }A^{\mu }=\pm M^{2}$ was originally proposed by
Dirac as a way to derive the electromagnetic current from the
additional excitations of the photon field, which now ceased to be
gauge degrees of freedom, avoiding in this way the problems
arising from considering point-like charges \cite{Diracfull}. As reported in %
\cite{Dirac1}, this theory requires the existence of an ether
emerging from quantum fluctuations, but not necessarily of the
violation of Lorentz invariance, due to an averaging process of the
quantum states producing such fluctuations.

In this work we explore the possibility of having a non-zero VEV of
the electromagnetic strength, so that we avoid the problem of
disturbing gauge invariance from the very beginning.
To this end we consider a non-linear version of Electrodynamics
with a potential term having an stable minimum
for a constant electromagnetic strength. The
paper is organized as follows: In Section II we make it plausible the existence of such gauge invariant potential $V_{eff}(F^2)$, arising as an effective low energy contribution from the integration of degrees of freedom corresponding to massive  gauge-bosons and fermions in a more fundamental theory. This potential also exhibits the right behavior in the low-intensity field approximation. Section III contains the construction
of the model starting from a non-linear version of Electrodynamics
which is spontaneously broken by choosing a vacuum characterized by
a constant electromagnetic tensor $C_{\mu \nu }$.
In Section IV we discuss the possible realizations of such vacuum
in terms of the associated electromagnetic fields $\mathbf{e}$ and
$\mathbf{b}$, identifying the Lorentz subgroups that remain
invariant after the breaking. The equations of motion corresponding
to the propagating sector, i.e. those arising from the quadratic
terms in the Lagrangian, are examined in Section V using a
covariant approach, as well as the standard $3+1$ decomposition.
Also, the modified Maxwell equations are presented. Section VI
contains a
summary of the modified photon dispersion relations $\omega =\omega (\mathbf{%
k})$, together with the expressions for the electromagnetic fields $\mathbf{E%
}$ and $\mathbf{B}$ of plane waves propagating in the different
vacua parameterized by $\mathbf{e}$ and $\mathbf{b}$. The model is shown to be stable
in the small Lorentz invariance violation approximation.
In section VII
we study the embedding\ of our model in the corresponding sector of
the Standard Model Extension \cite{KOSTELECKYED} and use the
experimentally established bounds upon the LIV parameters to
constrain the electromagnetic fields $\mathbf{e}$ and
$\mathbf{b}$. Finally we conclude with a summary of the
work together with some comments in Section VIII. The Appendix contains some details
of the calculation leading to the motivation described in Section II.

\section{Motivation}

One of the main challenges of the proposed approach is to make it
plausible
the existence of a gauge invariant potential $V(F_{\alpha \beta })\;$%
depending upon the electromagnetic tensor, which is stable for large
values of the electric and magnetic fields and which possess a
minimum for some constant value of $F_{\alpha \beta }$. In
particular, we will consider at this level only the case of magnetic
fields in order to avoid  further instabilities due to pair creation
in strong electric fields.

A first possibility that arises is to consider the
effective photon interaction in QED after integrating the
fermions. Unfortunately, a detailed calculation of this potential,
in the one-loop $\left(V_{EM}^{(1)}\right)$ and two-loop $\left(V_{EM}^{(2)}\right)$ approximations \cite{DR},
shows that it is
given by
\begin{equation}
V_{EM}(B)=\frac{1}{2}B^{2}+V_{EM}^{(1)}+V_{EM}^{(2)}=\frac{1}{2}B^{2}-\frac{%
e^{2}}{24\pi ^{2}}B^{2}\left[ \ln \left( \frac{eB}{m^{2}}\right) \right] -%
\frac{e^{4}}{128\pi ^{4}}B^{2}\ln \left( \frac{eB}{m^{2}}\right) ,
\label{EFFPOTQED}
\end{equation}%
in the limit \thinspace when
$B>>B_{0}=m^{2}/e$, which clearly is unbounded from below.

A new possibility arises in a further generalization of the
mechanism proposed in Refs. \cite{Bjorken, KT}, restricted to the
gauge invariant case. We start from a conventional gauge theory
including fermions, gauge fields and Higgs fields which provide
masses, via spontaneous symmetry breaking, to the gauge fields,
except for the photon potential $A_{\mu }$, which carries the
electromagnetic $U(1)$ gauge invariance. For simplicity we consider
only one fermion species, as in Ref.\cite{Bjorken}, but the idea can
be generalized to many of them. In the latter case, the subsequent integration over the
fermionic degrees of freedom could be  made only for those having a mass
larger than a given scale, thus leaving unintegrated the lightest
ones.

Integrating the massive gauge bosons of this theory leads to a
power series in bilinears in the fermion fields, which are
suppressed by powers of the energy scale $\Lambda $ that
characterizes the SSB mechanism providing the mass to the gauge
vector bosons \cite{KT}. We focus upon the neutral sector of the model where
the following contributions to the effective Lagrangian arise
\begin{equation}
\frac{1}{\Lambda ^{6n-4}}\left[ \sum_a \left( \bar{\Psi}M_{a}\Psi \right)
\left( \bar{\Psi}M^{a}\Psi \right) \right] ^{n},\;\;\frac{1}{\Lambda
^{8n-4}}\left[\sum_a \left( \bar{\Psi}M_{a}D_{\nu }\Psi \right) \left(
\bar{\Psi}M^{a}D^{\nu }\Psi \right) \right] ^{n}, \dots,
\end{equation}%
We are considering all possible contributions which are consistent
with gauge invariance, the remaining symmetry of this sector of the
theory. Notice also that terms containing the $U(1)$ covariant
derivative $D_{\nu }=\left(
\partial _{\nu }+ieA_{\nu }\right)$ are further suppressed with respect to
those without it. Also we consider only the $n=1$ case, which
provides a gauge invariant generalization of the model in Ref. \cite{Bjorken}. Thus
our starting point
is%
\begin{equation}
L_{eff}=\bar{\Psi}\left( i\gamma ^{\mu }\left( \partial _{\mu
}+ie{\tilde A}_{\mu }\right) -m\right) \Psi -\sum_{M,a}\frac{r_{M}}{\Lambda
^{2}}\left[ \left( \bar{\Psi}M_{a}\Psi \right) \left(
\bar{\Psi}M^{a}\Psi \right) \right] .
\end{equation}%
The index $M: S, V, T, PV, PS$ labels the tensorial objects that constitute the
Dirac basis: Scalar, Vector, Tensor, Pseudovector and Pseudoscalar, respectively.
The generic index $a$ labels the covariant (contravariant) components of each  tensor
class. More details are given in Table I of the Appendix. We follow the Dirac algebra conventions of
 Ref. \cite{BD}, with appropriate factors chosen in such a way that each current  $\left( \bar{%
\Psi}M^{a}\Psi \right)$ is real. In this way, the proposed
generalization includes all possible fermionic quartic
interactions. Also, the coefficients $r_{M}$ are assumed to be of
order one.

The fermions
can be integrated next by introducing the corresponding auxiliary fields $
(C_M)^{a}$ for each real current $(j_M)_{a}=\bar{\Psi}M_{a}\Psi $, as suggested in Ref. %
\cite{KT}, and following standard manipulations in the path integral
formulation. Some details are given in the Appendix.

Our main goal is to estimate the additional corrections provided by
such
currents to the purely electromagnetic effective potential (\ref{EFFPOTQED}%
) and to investigate whether or not it is possible to generate a net
positive contribution in the high-intensity field limit. To have a preliminary
simpler expression for the resulting extension we choose to consider
separately  each of the contributions arising from the different
currents and only add them to the total effective potential at the
end. In this way we are nor considering possible interference terms
among the currents, which nevertheless could be calculated within
the approximations performed. Calling $V^{(2)}$ the finite part of the total
effective potential arising from the currents $j_{M}$, given in Eq.(\ref{TOTALCONTR}) of the Appendix, we obtain
\begin{eqnarray}
B &\rightarrow &\infty :\;\;\;V^{(2)}=\left( \frac{1}{16\pi ^{2}}%
\right) ^{2}\;\left( \frac{m}{\Lambda }\right) ^{2}e^{2}B^{2}\left[
\ln \left( \frac{2eB}{m^{2}}\right) \right]
^{2}\sum_{M}r_{M}\,\theta_{M},
\label{finalinf} \\
B &\rightarrow &0:\;\;\;V^{(2)}=\left( \frac{1}{16\pi
^{2}}\right)
^{2}\left( \frac{m}{\Lambda }\right) ^{2}\left( \frac{eB^{2}}{3m^{2}}+\frac{1%
}{18}\frac{e^{2}B^{4}}{m^{4}}\right)^2 \sum_{M}r_{M}\,\theta_{M}.
\label{finalcero}
\end{eqnarray}
Here $m$ is the mass of the integrated fermion which is such that
$m>\Lambda$. The main point of the estimation is that we obtain
\begin{equation}
\Theta =\sum_{M}r_{M}\, \theta _{M}=4r_{S}+16r_{V}+48r_{T}-16r_{PV
}-4r_{PS},
\end{equation}%
which can be made positive from a judicious choice of the parameters $%
r_{M}$.

In the high-intensity field approximation, the dominant term in the
electromagnetic contribution goes like
\begin{equation}
V_{EM}(B)\simeq -e^{2}B^{2}\ln \left( \frac{eB}{m^{2}}\right) ,
\end{equation}%
which can be overcame by the contribution from the additional currents $\;$%
\begin{equation}
V^{(2)}\simeq\;\left( \frac{m}{\Lambda }\right)
^{2}e^{2}B^{2}\left[ \ln \left( \frac{2eB}{m^{2}}\right) \right]
^{2}\Theta,
\end{equation}%
provided $\Theta >0.$

As we mention in the Appendix, this preliminary estimation has only considered an
appropriate regularization of the divergent integrals, but is
missing an adequate renormalization of the total effective
Lagrangian $L_{eff}$. Our
renormalization conditions would be%
\begin{equation}
\lim_{B\rightarrow 0}L_{eff}=\rho B^2,\qquad
\rho >0,
\end{equation}%
in such a way that the effective potential is a decreasing function
in the vicinity of $B=0.$ The final normalization to the value of $
L_{eff}=-b^{2}/2$ will be imposed at the end of the procedure, after
we have expanded the magnetic field around an stable minimum $B_{Min}$ of
$V_{eff}$, such that
\begin{equation}
B=B_{Min}+b
\end{equation}%
and we are left with the physical component $b$. The renormalization
procedure should be analogous to the one carried in Ref. \cite{DR}\
for the two-loop contribution $L_{EM}^{(2)}(A)$ to the effective electromagnetic action and it is rather involved due
to the complicated field dependence of the regularized version of
the divergent integrals. This calculation is beyond the scope of
these preliminary estimations.

Having an effective potential with the characteristics described
above guarantees the existence of an absolute minimum, which we
assume to arise at values lower that the critical magnetic and
electric fields, in such a way to avoid  fermion pair production in
the later case. We also generalize this idea to the case of an
arbitrary constant electromagnetic field $F_{\mu \nu }$. The study of the  stability of the model under radiative corrections is beyond the scope of the present work.

\section{The Model}

In the previous section we have made it plausible the existence of a gauge invariant potential $V_{eff}(F^2)$ having an stable minimum and arising as an effective low energy model from  the integration of degrees of freedom corresponding to massive  gauge-bosons and fermions in a more fundamental theory. This potential also exhibits the right behavior in the low-intensity field approximation. To examine the  dynamical consequences of the symmetry breaking  we have to adopt a convenient parametrization of  such potential. There are various choices in the literature which would translate into the following for  our case \cite{choice}
\begin{equation}
V(F)= \lambda(F^2 \pm C^2), \qquad V(F)= \frac{\lambda}{4}(F^2 \pm C^2)^2, \qquad
\end{equation}
where $C_{\mu\nu}$  is the constant value of the electromagnetic field at the minimum. Instead we will adopt the standard Ginzburg-Landau parametrization
\begin{equation}
V(F_{\mu \nu })=\frac{1}{2}\alpha F^{2}+\frac{\beta }{4}\left(
F^{2}\right) ^{2}.  \label{POT}
\end{equation}%
 Since the  basic variable we want to deal with is $F_{\mu\nu}$, we start from the effective Lagrangian
\begin{equation}
L(F_{\alpha \beta },X_{\mu })=-V(F_{\alpha \beta })-\bar{F}^{\nu \mu
}\partial _{\nu }X_{\mu }, \quad F_{\alpha \beta }=-F_{\beta \alpha
}, \label{LAGINI}
\end{equation}%
where the dual field $\bar{F}^{\nu \mu }$ is
\begin{equation}
\bar{F}^{\nu \mu }=\frac{1}{2}\epsilon ^{\nu \mu \alpha \beta
}F_{\alpha \beta }
\end{equation}%
and the $X_{\mu }$ are just Lagrange multipliers that will finally
impose the condition that the field strength satisfies the
Bianchi identity and can then be expressed in terms of the vector potential. The above Lagrangian is similar to that used in
the standard formulation of non-linear electrodynamics
\cite{PLEBANSKI}. Our conventions are $ \eta _{\mu \nu }=diag(+,-,-,-),\quad \epsilon
^{0123}=+1,\quad \epsilon _{123}=+1$.

The next step is to determine the vacuum configuration of the
theory,
corresponding to the minimum of the energy $E$. The energy density $T^{00}\;$%
is calculated via the standard Noether theorem starting from the Lagrangian (%
\ref{LAGINI})\ and produces
\begin{equation}
E=\int d^{3}x\;T^{00}=\int d^{3}x\;\left[ V(F_{\alpha \beta })+\bar{F}%
^{i\alpha }\left( \partial _{i}X_{\alpha }\right) \right] .
\end{equation}%
The extremum conditions, obtained by varying \ the independent fields $%
F_{\alpha \beta }\;$and $X_{\alpha }$, are$\;$%
\begin{eqnarray}
\frac{\delta E}{\delta F_{\rho \sigma }} &=&\frac{\partial
V}{\partial F_{\rho \sigma }}+\left( \partial _{i}X_{\alpha }\right)
\epsilon ^{i\alpha
\rho \sigma }=0,  \label{EXTRE1} \\
\frac{\delta E}{\delta X_{\alpha }} &=&-\frac{1}{2}\epsilon
^{i\alpha \rho \sigma }\partial _{i}F_{\rho \sigma }=0.
\label{EXTRE2}
\end{eqnarray}

In order to maintain four dimensional translational invariance we
look for extrema of the fields which are independent of the
coordinates. From Eqs. (\ref{EXTRE1}) and (\ref{EXTRE2})
we verify that this is possible provided the condition
\begin{equation}
\left( \frac{\partial V}{\partial F_{\rho \sigma }}\right)
_{Extr.}=0.
\end{equation}%
holds.
In other words, to find the vacuum configuration we have to
extremize the effective action, subjected to the condition that
$F_{\alpha \beta }$ and $
X_{\alpha }$ are constant fields. Applying these requirements to (\ref%
{LAGINI})\ plus the choice (\ref{POT}) we obtain
\begin{equation}
\frac{\partial V}{\partial F^{\mu \nu }}=0=\left( \alpha +\beta
F^{2}\right) F_{\mu \nu },
\end{equation}%
which is solved by a constant $\left( F_{Extr}\right) _{\alpha \beta
}\equiv C_{\alpha \beta }$ such that
\begin{equation}
\left( F^{2}\right) _{E}=-\frac{\alpha }{\beta }=C^{2}\neq 0.
\label{EXC1}
\end{equation}%
The expansion around the minimum $(C_{\mu \nu },\;C_{\mu })$ is
written as
\begin{equation}
F_{\alpha \beta }(x)= C_{\alpha \beta }+a_{\alpha \beta }(x), \qquad
X_{\mu
} = C_{\mu }+\bar{X}_{\mu }.  \label{EXP1} \\
\end{equation}%
Introducing such expansion in the potential (\ref{POT}), we arrive
to
\begin{eqnarray}
V(F_{\alpha \beta }) &=&V(C_{\alpha \beta })+\beta \left( C_{\alpha
\beta }a^{\alpha \gamma }\right) ^{2}+\beta \left( C_{\alpha \beta
}a^{\alpha
\beta }\right) \left( a_{\mu \nu }a^{\mu \nu }\right) +\frac{\beta }{4}%
\left( a_{\alpha \beta }a^{\alpha \beta }\right) ^{2},  \label{EXPOT1} \\
&\equiv &V(C_{\alpha \beta })+\bar{V}(a_{\alpha \beta }).
\label{EXPOT2}
\end{eqnarray}%
We notice that (\ref{EXPOT1})\ has no linear term in $a^{\alpha
\gamma }$, and also that the quadratic term $\beta \left( C_{\alpha
\beta }a^{\alpha \gamma }\right) ^{2}\;\;$is positive provided
\begin{equation}
\beta > 0,  \label{EXC2}
\end{equation}%
thus verifying the expansion around a minimum.\ On the other hand,
according to Eq. (\ref{EXC1}), the sign of $\alpha$ is arbitrary, in
such a way that it is opposite to the sign of $C^{2}$.

In this way, the spontaneously symmetry broken action is
\begin{equation}
S(a_{\alpha \beta },\bar{X}_{\mu }\;)=-\int d^{4}x\left(
\frac{1}{2}\epsilon
^{\nu \mu \alpha \beta }a_{\alpha \beta }\partial _{\nu }\bar{X}_{\mu }+\bar{%
V}(a_{\alpha \beta })+V(C)\right) ,  \label{LAG2}
\end{equation}%
where we have eliminated the total derivatives arising from the
shifts of the fields (\ref{EXP1}) performed in the original
Lagrangian (\ref{LAGINI}).

Next we consider the equations of motion derived from (\ref{LAG2})\
and show that the Lagrange multiplier $\bar{X}_{\mu }\;$is fully
determined up to gauge transformation $\bar{X}_{\mu }\longrightarrow
\bar{X}_{\mu }+\partial _{\mu }\chi $.\ The equations are
\begin{eqnarray}
\delta a_{\alpha \beta } &:&\;-\epsilon ^{\nu \mu \alpha \beta
}\partial _{\nu }\bar{X}_{\mu }-2\frac{\partial \bar{V}}{\partial
a_{\alpha \beta }}=0,
\label{EQMOT1} \\
\delta \bar{X}_{\mu } &:&\;\;\;\epsilon ^{\nu \mu \alpha \beta
}\partial _{\nu }a_{\alpha \beta }=0.  \label{EQMOT2}
\end{eqnarray}%
Eq. (\ref{EQMOT2}) establishes that the two form $a$ is closed, i.e.
$da=0.$ The Hodge-De Rham theorem says that the most general
solution to this is obtained by requiring
\begin{equation}
a=dA+ls,  \label{CAMPO}
\end{equation}%
where $A$ is a one form, $l$ is a constant and $s$ is an harmonic
two-form. \ Since in our case we naturally have one such form at our
disposal, arising precisely from the chosen vacuum, we take
\begin{equation}
s=\frac{1}{2}C_{\mu \nu }dx^{\mu }dx^{\nu},
\end{equation}
which is certainly harmonic because it is constant. Calling $dA=f$
\ we
have%
\begin{equation}
a_{\alpha \beta }(x)=lC_{\alpha \beta }+f_{\alpha \beta }(x).
\label{CAMPO1}
\end{equation}%
From (\ref{EQMOT1})\ we obtain
\begin{equation}
\partial _{\alpha }\left( \frac{\partial \bar{V}}{\partial a_{\alpha \beta }}%
\right) =0  \label{EQMOT3}
\end{equation}%
which are the generalization of Maxwell equations for the potential $%
A_{\alpha }$.
From (\ref{EQMOT1}) we obtain $\partial _{\nu
}\bar{X}_{\mu
}-\partial _{\mu }\bar{X}_{\nu }$ yielding%
\begin{equation}
\left( \partial _{\rho }\bar{X}_{\sigma }-\partial _{\sigma
}\bar{X}_{\rho
}\right) =\epsilon _{\rho \sigma \alpha \beta }\frac{\partial \bar{V}}{%
\partial a_{\alpha \beta }}\equiv R_{\rho \sigma }(A_{\beta }).
\label{EQMOT4}
\end{equation}%
We verify that $R_{\rho \sigma }(A_{\beta })$ can in fact be
written as the left hand side of Eq. (\ref{EQMOT4}) by calculating $\epsilon^{\alpha \beta \rho
\sigma }\partial _{\beta }R_{\rho \sigma }$. The result is zero in
virtue of the equations of motion
for $A_{\beta }$. Fixing the corresponding gauge freedom by setting $\partial^{\mu }\bar{X}%
_{\mu }=0$, for example, we obtain
\begin{equation}
\partial ^{2}\bar{X}_{\sigma }=\partial _{\rho }R_{\rho \sigma }(A_{\beta }),
\label{EQMOT5}
\end{equation}%
which completely determines the Lagrange multiplier $\bar{X}_{\sigma }.\;$%
Going back to the action (\ref{LAG2}), integrating by parts and
substituting the constraint (\ref{EQMOT2}) we are left with the
reduced effective action
\begin{equation}
S(A_{\alpha })=-\int d^{4}x\;\left( \bar{V}(a_{\alpha \beta
})+V(C)\right) ,\;\;\;\;\;\;\;a_{\alpha \beta }(x)=lC_{\alpha \beta
}+\partial _{\alpha }A_{\beta }-\partial _{\beta }A_{\alpha }.\;\;
\label{ACT3}
\end{equation}%
The resulting equations of motion are
\begin{equation}
\delta A_{\alpha }:\;\;\partial _{\beta }\left( \frac{\partial \bar{V}}{%
\partial a_{\beta \alpha }}\right) =0,
\end{equation}%
which coincide with the previous ones in Eq. (\ref{EQMOT3}).

It is convenient to rename the dimensionless parameter $l=\xi -1$.
Substituting (\ref{CAMPO1}) in (\ref{ACT3}) and defining
\begin{equation}
C_{\mu \nu }=\frac{1}{2\xi }D_{\mu \nu },\hspace{0.75em}\hspace{0.75em}%
\hspace{0.75em}\hspace{0.75em}{\cal B}=\frac{\beta }{4}>0,  \label{COND2}
\end{equation}%
where%
\begin{equation}
\beta =\frac{1}{2\left( \xi ^{2}-1\right) \left( C^{\alpha \beta
}C_{\alpha \beta }\right) },\hspace{0.75em}\hspace{0.75em}\left( \xi
^{2}-1\right) \left( C^{\alpha \beta }C_{\alpha \beta }\right) >0,
\label{COND1}
\end{equation}%
we obtain our final effective action
\begin{equation}
S(A_{\alpha })=\int d^{4}x\left( -\frac{1}{4}\left[1-D^{2}{\cal B}\right] D^{2}-%
\frac{1}{4}f_{\mu \nu }f^{\mu \nu }-{\cal B}\left[ \left( D_{\mu \nu
}f^{\mu \nu }\right) +\left( f_{\mu \nu }f^{\mu \nu }\right) \right]
^{2}\right), \label{LAGFIN1}
\end{equation}%
where \ $f_{\alpha \beta }(x)=\partial _{\alpha }A_{\beta }-\partial
_{\beta }A_{\alpha }$.  We recognize the standard
Maxwell kinetic term
in the right hand side  of Eq.(\ref{LAGFIN1}). The only restriction now is ${\cal B}>0$, with  $D^{2}$ arbitrary.

Let us emphasize that the case of purely spontaneously broken
Lorentz invariance really corresponds to the singular choice $\xi
=1,(l=0)$.$\;$In this case, the corresponding action would not
possess the standard kinetic term $-\frac{1}{4}f_{\mu \nu }f^{\mu
\nu }\;$but will start with the quadratic term\textbf{\ }$\left(
D_{\mu \nu }f^{\mu \nu }\right) ^{2}$. This very interesting case is
not considered here and its investigation is postponed for future
work.


\section{Symmetry algebras arising from different choices of the vacuum}

In this section we study the possible vacua allowed by the tensor symmetry
breaking and we also identify the corresponding subgroups of the Lorentz group
which are left invariant after the breaking. In \ order to do this, it is
convenient to parameterize the background field $D_{\mu \nu }$ in terms of \
three dimensional components
\begin{equation}
D_{ij}=-\epsilon _{ijm}b_{m},\hspace{0.75em}\hspace{0.75em}\hspace{0.75em}%
D_{0i}=e_{i},\hspace{0.75em}\epsilon _{123}=1\hspace{0.75em}\hspace{0.75em}%
\hspace{0.75em}\hspace{0.75em}
\end{equation}%
\begin{equation}
\left[ D_{\mu \nu }\right] =\left[
\begin{array}{cccc}
0 & e_{1} & e_{2} & e_{3} \\
-e_{1} & 0 & -b_{3} & b_{2} \\
-e_{2} & b_{3} & 0 & -b_{1} \\
-e_{3} & -b_{2} & b_{1} & 0%
\end{array}%
\right]
\end{equation}%
which will mix when going to another reference frame via a passive Lorentz
transformation. Since we have two vectors that determine a plane we choose a
coordinate frame where
\begin{equation}
\mathbf{b=(}0,0,b\mathbf{),\hspace{0.75em}\hspace{0.75em}e\mathbf{=}}(0,%
\hspace{0.75em}e_{2}=|\mathbf{e|}\sin \psi ,\hspace{0.75em}e_{3}=|\mathbf{e|}%
\cos \psi ),\hspace{0.75em}  \label{PREFCOORD}
\end{equation}%
That is to say, we have chosen the plane of the two vectors as the $\left(
z-y\right) $ plane, with the vector $\mathbf{b\hspace{0.75em}}$defining the $%
z$-direction and $\psi \hspace{0.75em}$being the angle between $\mathbf{b%
\hspace{0.75em}}$and $\mathbf{e}$. In this way the matrix representing the
vacuum is

\begin{equation}
\left[ D_{\mu \nu }\right] =\left[
\begin{array}{cccc}
0 & 0 & e_{2} & e_{3} \\
0 & 0 & -b & 0 \\
-e_{2} & b & 0 & 0 \\
-e_{3} & 0 & 0 & 0%
\end{array}%
\right] .
\end{equation}

The most general infinitesimal generator \ $G$, including \ Lorentz
transformations plus dilatations is
\begin{equation}
G=\left[ G_{\hspace{0.75em}\nu }^{\mu }\right] =-i\left[
\begin{array}{cccc}
z & x_{1} & x_{2} & x_{3} \\
x_{1} & z & -y_{3} & y_{2} \\
x_{2} & y_{3} & z & -y_{1} \\
x_{3} & -y_{2} & y_{1} & z%
\end{array}%
\right] .  \label{GEN1}
\end{equation}%
Motivated by the work in Ref. \cite{HL} we are including conformal
dilatations $D$ among our generators. Within this restricted Poincare
algebra, this generator commutes with the remaining ones corresponding to
pure Lorentz transformations and can be realized as a multiple of the
identity. We do this in order to explore the possibility of having the
largest possible invariant sub-algebra after the symmetry breaking.

The condition for the vacuum to be invariant under the infinitesimal
transformations is
\begin{equation}
0=G_{\hspace{0.75em}\alpha }^{\mu }D^{\alpha \nu }+G_{\;\alpha }^{\nu
}\;D^{\mu \alpha }  \label{COND3}
\end{equation}%
and the resulting equations are:
\begin{eqnarray}
x_{2}b-y_{2}e_{3} &=&-y_{3}e_{2},\;\;\;x_{2}e_{3}+y_{2}b=x_{3}e_{2}, \\
x_{1}e_{2}+2zb &=&0,\;\;\;y_{1}e_{2}+2ze_{3}=0, \\
-x_{1}b+y_{1}e_{3} &=&2ze_{2},\;\ x_{1}e_{3}+y_{1}b=0.
\end{eqnarray}%
Next we consider the solutions of the above system of
equations corresponding to the seven non-trivial possibilities dictated by
the choices in the arrangement $(b,\;e_{2},\;e_{3})$ in the above coordinate
system. In the sequel we use the notation of Ref.\cite{KAKU} for the
Lorentz group generators.
\subsection{ Case $\;b=e_{3}=0,e_{2}\neq 0$}
This leads to
\begin{equation}
z=0,\;\;x_{1}=y_{1}=0,\;\;\;\;\;\;x_{3}=y_{3}=0.
\end{equation}%
We have two free parameters, $x_{2},y_{2}$ and the generator is
\begin{equation}
G=x_{2}\;K^{2}-y_{2}\;J^{2}.
\end{equation}
\subsection{Case $b=0,\;e_{3}\neq 0,\;\;e_{2}=0$}
Here we have
\begin{equation}
z=0,\;\;\;x_{1}=y_{1}=0,\;\;\;x_{2}=y_{2}=0,\;\;\;\;\;\;
\end{equation}%
so that we are left with a two parameter $(x_{3},\;y_{3})$ subalgebra with
generator
\begin{equation}
G=x_{3}K^{3}-y_{3}J^{3}.
\end{equation}
\subsection{Case$\;b\neq 0,e_{3}=0,\;\;e_{2}=0$}
This case is analogous to the previous one. We have%
\begin{equation}
z=0,\;\;\;x_{1}=y_{1}=0,\;x_{2}=y_{2}=0,\;\;\;\;\;
\end{equation}%
with a two parameter $(x_{3},\;y_{3})\;$subalgebra.
\subsection{Case$\;\;b=0\mathbf{,\;}e_{3}\neq 0,\;e_{2}\neq 0$}
In this case we have
\begin{equation}
z(e_{2}^{2}+e_{2}^{2})\ =0\;\;\rightarrow z=0,
\end{equation}%
\begin{equation}
x_{1}=0,\;\;y_{1}=0,\;\;y_{2}=y_{3}\frac{e_{2}}{e_{3}},\;\;x_{2}=x_{3}\frac{%
e_{2}}{e_{3}},
\end{equation}%
yielding a two parameter $(x_{3},y_{3})$\ subalgebra with generator%
\begin{equation}
G=x_{3}\left( \frac{e_{2}}{e_{3}}K^{2}+K^{3}\right) -y_{3}\left( J^{3}+\frac{%
e_{2}}{e_{3}}J^{2}\right).
\end{equation}%
Calling
\begin{equation}
G_{x_{3}}=\frac{e_{2}}{e_{3}}K^{2}+K^{3},\;\;G_{y_{3}}=-\;\left( J^{3}+\frac{%
e_{2}}{e_{3}}J^{2}\right),
\end{equation}%
we can show that
\begin{equation}
\left[ G_{x_{3}},\;G_{y_{3}}\right] =0.
\end{equation}
\subsection{Case $\;b\neq 0,\;e_{2}\neq 0,\;e_{3}=0$}
Here we have%
\begin{equation}
y_{3}=-x_{2}\frac{b}{e_{2}},\;\;\;x_{3}=+y_{2}\frac{b}{e_{2}}\;,\;y_{1}=0
\end{equation}%
\ \ and the consistency condition
\begin{equation}
z(b^{2}-e_{2}^{2})=0,\;\;\;
\end{equation}%
which leads to two possibilities
\subsubsection{Subcase $\;z=0$}
Here we have
\begin{equation}
y_{3}=-x_{2}\frac{b}{e_{2}},\;\;\;x_{3}=+y_{2}\frac{b}{e_{2}}%
\;,\;y_{1}=0,\;x_{1}=0
\end{equation}%
and we are left with a two-parameter subgroup where%
\begin{equation}
G=x_{2}\left( K^{2}+\frac{b}{e_{2}}J^{3}\right) +y_{2}\left( \frac{b}{e_{2}}%
K^{3}-J^{2}\right).
\end{equation}%
Calling%
\begin{equation}
G_{x_{2}}=\left( K^{2}+\frac{b}{e_{2}}J^{3}\right) ,\;\;\;\;G_{y_{2}}=\left(
\frac{b}{e_{2}}K^{3}-J^{2}\right),
\end{equation}%
we can verify that%
\begin{equation}
\left[ G_{x_{2}},\;\;G_{y_{2}}\right] =0.
\end{equation}
\subsubsection{Subcase $z\neq 0$}
Here we have
\begin{equation}
b^{2}-e_{2}^{2}=0\rightarrow b=se_{2},\;\;s=\pm 1,  \label{NULLC}
\end{equation}%
which corresponds to a plane wave VEV.

Summarizing we have
\begin{equation}
y_{1}=0,\;\;\;\;\;x_{1}=-2sz,\;\;\;x_{2}=-sy_{3},\;y_{2}=sx_{3},
\end{equation}%
Here we are left with a three parameter Lie algebra $\left(
z,\;x_{3},\;y_{3}\right) $\ and the generator is
\begin{equation}
G=-z\left( 2sK^{1}+iI\right) +x_{3}\left( K^{3}-sJ^{2}\right) -y_{3}\left(
sK^{2}+J^{3}\right).
\end{equation}
Defining
\begin{equation}
G_{z}=-\left( 2sK^{1}+iI\right) ,\;\;\;G_{x}=K^{3}-sJ^{2},\;\;G_{y}=-\left(
J^{3}+sK^{2}\right),
\end{equation}%
we obtain the algebra
\begin{equation}
\left[ G_{z},G_{x}\right] =2iG_{x},\;\;\left[ G_{z},G_{y}\right]
=2iG_{y},\;\;\;\;\left[ G_{x},G_{y}\right] =0,
\end{equation}%
which is isomorphic to HOM(2).

The condition (\ref{NULLC}) implies that $D^{2}=D_{\mu \nu }D^{\mu \nu }=0$.
Since ${\cal B}$ defined in Eq. (\ref{COND2}) has to be finite and positive we
have to be careful in the limiting procedure leading to this quantity. Also
we need to have $C_{\mu \nu }C^{\mu \nu }\neq 0$, in order that the
original parameters $\alpha $ and $\beta $ are well defined.\ Let us
consider \
\begin{equation}
e_{2}=sb+\epsilon ,\;e_{3}=0,\;\epsilon \rightarrow 0\;,\;\;
\end{equation}%
such that%
\begin{equation}
D^{2}=2(b^{2}-e_{2}^{2})=-4sb\epsilon .
\end{equation}%
The choice
\begin{equation}
\xi =\gamma \sqrt{\epsilon },\;
\end{equation}%
with $\gamma$ arbitrary, guarantees that%
\begin{equation}
{\cal {\cal B}}=\frac{\gamma ^{2}}{8sb}
\end{equation}%
is finite. Besides we must demand that%
\begin{equation}
sb>0.
\end{equation}%
Next we consider%
\begin{equation}
C_{\mu \nu }=\frac{1}{2\gamma \sqrt{\epsilon }}D_{\mu \nu },\;\;\;\;C^{2}=-%
\frac{4sb\epsilon }{4\gamma ^{2}\epsilon }=-\frac{sb}{\gamma ^{2}}<0,
\end{equation}%
which ensures that $C^{2}\;$is finite. The fact that $C^{2}\;$is negative is
consistent with the second condition in (\ref{COND1}). Summarizing, in this
case we have%
\begin{equation}
\beta =\frac{\gamma ^{2}}{2sb}>0,\;\;\;\;\alpha =\frac{1}{2},
\end{equation}%
showing that the limiting procedure is well defined.
\subsection{Case $b\neq 0,\;\;\;e_{3}\neq 0,\; e_{2}=0$}
We have%
\begin{equation}
\;z=0,\qquad  x_{1}=y_{1}=0,\qquad  x_{2}=y_{2}=0.
\end{equation}%
The free parameters here  are $x_{3},y_{3}.$
\subsection{Case $\;b\neq 0,\;e_{2}\neq 0,\;e_{3}\neq 0\;\;\;$}
Here
\begin{equation}
y_{3}=-x_{2}\frac{b}{e_{2}}+y_{2}\frac{e_{3}}{e_{2}},\;\;\;x_{3}=x_{2}\frac{%
e_{3}}{e_{2}}+y_{2}\frac{b}{e_{2}}\;,
\end{equation}%
together with  the consistency condition
\begin{equation}
z(b^{2}-e_{3}^{2}-e_{2}^{2})=0,\;\;\;\;zbe_{3}=0,
\end{equation}%
which leads to $z=0$. We have two free parameters: \ $x_{2},y_{2}$ . The
generators are%
\begin{equation}
G=x_{2}\left( K^{2}+\frac{e_{3}}{e_{2}}K^{3}+\frac{b}{e_{2}}J^{3}\right)
-y_{2}\left( J^{2}-\frac{b}{e_{2}}K^{3}+\frac{e_{3}}{e_{2}}J^{3}\right).
\end{equation}%
Defining
\begin{equation}
G_{x_{2}}^{\prime }=\left( K^{2}+\frac{e_{3}}{e_{2}}K^{3}+\frac{b}{e_{2}}%
J^{3}\right) ,\;\;\;G_{y_{2}}^{\prime }=-\left( J^{2}-\frac{b}{e_{2}}K^{3}+%
\frac{e_{3}}{e_{2}}J^{3}\right),
\end{equation}%
we can verify that
\begin{equation}
\left[ G_{x_{2}}^{\prime },\;G_{y_{2}}^{\prime }\right] =0.
\end{equation}%
Summarizing, all the  two-parameter subalgebras that leave the
vacuum invariant are isomorphic to $T(2)$, while the only three-parameter
subalgebra, corresponding to the case (E-2), is isomorphic to $HOM(2)$%
\textbf{.}

\subsection{THE SYMMETRIES OF THE ACTION}

To conclude this Section, let us consider the full effective action%
\begin{equation}
L=-\frac{1}{4}f_{\mu \nu }f^{\mu \nu }-{\cal B}\left[ \left( D^{\mu \nu }f_{\mu \nu
}\right) +\left( f_{\mu \nu }f^{\mu \nu }\right) \right] ^{2}
\label{LAGFIN3}
\end{equation}%
and verify the invariance under the Lorentz plus \ scale transformations
that leave the vacuum invariant given by the generator (\ref{GEN1})\
together with the condition (\ref{COND3}).\ The term$\;f_{\mu \nu }f^{\mu
\nu }\;$is invariant under the full Lorentz group plus scale
transformations, which includes also the second term in the parenthesis
proportional to ${\cal B}$. Let us verify the invariance of the term$\;\left(
D^{\mu \nu }f_{\mu \nu }\right)$ under the transformation (\ref{COND3}). We have
\begin{equation}
\delta \left( D^{\mu \nu }f_{\mu \nu }\right) =\left( \delta D^{\mu \nu
}\right) f_{\mu \nu } +\left( D^{\mu \nu }\delta f_{\mu \nu }\right) =\left(
D^{\mu \nu }\delta f_{\mu \nu }\right),
\end{equation}%
because (\ref{COND3})\ is the subset of transformations that leave the vacuum%
$\;D^{\mu \nu }$ invariant. The remaining transformations are given by those
of a contravariant tensor
\begin{equation}
\left( D^{\mu \nu }\delta f_{\mu \nu }\right) =-D^{\mu \nu }\left( f_{\alpha
\nu }G_{\;\;\mu }^{\alpha }+f_{\mu \alpha }G_{\;\;\nu }^{\alpha }\right)
=-f_{\rho \sigma }\left( G_{\;\;\alpha }^{\rho }D^{\alpha \sigma
}+G_{\;\;\alpha }^{\sigma }D^{\rho \alpha }\right) =0
\end{equation}%
because the last equation in the right hand side just reproduces the vacuum invariance
condition (\ref{COND3}). In these way, the symmetries of (\ref{LAGFIN3}) are
in fact those of the vacuum $D^{\mu \nu }.$

\section{The equations of motion: the propagating sector}

To study the propagation properties we consider only the quadratic terms in
the effective Lagrangian%
\begin{equation}
L_{0}=-\frac{1}{4}f_{\mu \nu }f^{\mu \nu }-{\cal B}\left( f_{\mu \nu }D^{\mu \nu
}\right) ^{2},\;\;\;
\end{equation}%
where we recall that $f_{\alpha \beta }=\partial _{\alpha }A_{\beta
}-\partial _{\beta }A_{\alpha }$. The equations are
\begin{equation}
\left( \partial ^{2}A_{\beta }-\partial _{\beta }\partial ^{\alpha
}A_{\alpha }\right) =-8{\cal B}D_{\alpha \beta }\partial ^{\alpha }\left( D^{\mu
\nu }\partial _{\mu }A_{\nu }\right).  \label{EQMOT}
\end{equation}%
We have verified the consistency of the above when taking $\partial ^{\beta
} $.

It is convenient to define
\begin{equation}
X=D^{\mu \nu }\partial _{\mu }A_{\nu }
\end{equation}%
and to introduce the notation
\begin{eqnarray}
D_{\alpha k}\partial ^{\alpha } &=&D_{k}=\left[ D_{0k}\partial
_{0}-D_{lk}\partial _{l}\right],  \label{P1} \\
D_{0} &=&D_{i0}\partial _{i}.  \label{P2}
\end{eqnarray}%
In this way we have
\begin{equation}
X=-\left( D_{j}A_{j}+D_{0}A_{0}\right) ,
\end{equation}%
together with%
\begin{equation}
D_{0}\partial _{0}=-\partial _{l}D_{l}.
\end{equation}
\subsection{Covariant formulation}
The modified dispersion relations are found very easily by manipulating
Eqs.(\ref{EQMOT}). In momentum space
\begin{equation}
A_{\mu }(x)=\int d^{4}x\;A_{\mu }(k)e^{-ik_{\mu }x^{\mu }}
\end{equation}%
and choosing the Lorentz gauge, these equations reduce to
\begin{eqnarray}
k^{2}A_{\mu }+2p_{\mu }(p^{\nu }A_{\nu })=0 &&,\;\;\;p^{\alpha }\equiv 2%
\sqrt{{\cal B}}D^{\alpha \beta }k_{\beta },  \label{COV} \\
k^{\nu }A_{\nu }=0 &&.  \label{LORG}
\end{eqnarray}%
The vector $p^{\alpha }\;$is proportional to the momentum space
version of the vector $D_{\mu }\;$introduced in Eqs. (\ref{P1}) and
(\ref{P2}).

Multiplying the first relation in (\ref{COV}) by $p^{\mu }$, it follows
that
\begin{equation}
(k^{2}+2p^{2})(p^{\nu }A_{\nu })=0.  \label{DR1}
\end{equation}%
Moreover\ $p^{\nu }A_{\nu }\;$is gauge invariant. In fact, in coordinate
space is proportional to $D^{\alpha \beta }f_{\alpha \beta }$ . So this
component is physical and has a dispersion relation given by
\begin{equation}
k^{2}+2p^{2}=0.
\end{equation}%
If $(k^{2}+2p^{2})$ is not zero in Eq.(\ref{DR1}), it follows that $
p^{\nu }A_{\nu }=0$ . In four dimensions, this condition plus the Lorentz
gauge leaves two degrees of freedom. But the Lorentz gauge permits a further
gauge transformation with parameter $\lambda $ such that
\begin{equation}
\partial ^{2}\lambda =0,
\end{equation}%
which leaves only one degree of freedom as it should be. Moreover from the
first equation in (\ref{COV}), this remaining degree of freedom satisfies
\begin{equation}
k^{2}A_{\mu }=0,
\end{equation}%
so its dispersion relation is
\begin{equation}
k^{2}=0.  \label{DR2}
\end{equation}%
The general solution of (\ref{DR1}) is
\begin{equation}
p^{\nu }A_{\nu }=-\lambda _{1}(k)\delta (k^{2}+2p^{2}).
\end{equation}%
Putting it back into (\ref{COV}), we get
\begin{equation}
A_{\mu }=a_{\mu }(k)\delta (k^{2})+2\frac{\lambda _{1}(k)}{k^{2}}\delta
(k^{2}+2p^{2})p_{\mu },\;\;\;\;\;a_{\mu }(k)k^{\mu }=0, a_{\mu} ( k )
p^{\mu} = 0.  \label{sol}
\end{equation}%
From (\ref{sol}), we obtain the electromagnetic tensor
\begin{equation}
f_{\mu \nu }=(k_{\mu }a_{\nu }(k)-k_{\nu }a_{\mu }(k))\delta (k^{2})+2\frac{%
\lambda _{1}(k)}{k^{2}}\delta (k^{2}+2p^{2})(k_{\mu }p_{\nu }-k_{\nu }p_{\mu
}).
\end{equation}
It represents a plane wave with dispersion relation $k^{2}=0$ (the magnetic
and electric fields are the standard ones, perpendicular to $\mathbf{k}$), \
plus a plane wave propagating in the direction $\mathbf{k}$ with dispersion
relation $k^{2}+2p^{2}=0$.

The fields for the second wave are
\begin{equation}
E_{j}=f_{0j}=2(k_{0}p_{j}-k_{j}p_{0})\frac{\lambda _{1}(k)}{k^{2}},%
\qquad B_{j}=2\epsilon _{jkl}(k_{k}p_{l})\frac{\lambda _{1}(k)}{%
k^{2}}.  \label{COVFIELDS}
\end{equation}

Notice that $\mathbf{E}$ is perpendicular to $\mathbf{B}$, $\mathbf{B}$%
\textbf{\ }is perpendicular to $\mathbf{k}$, \ but $\mathbf{E}$ is not
necessarily orthogonal to $\mathbf{k}$. Moreover, this wave exists only if $%
k^{2}\neq 0$.
\subsection{3+1 Decomposition \qquad}
An alternative way to determine the polarization of the plane wave solutions
is by splitting Eq. (\ref{EQMOT}) into the components $\;0,i,\;$which yields
\begin{equation}
\left( -\nabla ^{2}+2bD_{0}^{2}\right) A_{0}-\partial _{0}\partial
^{i}A_{i}=-8{\cal B}D_{0}D_{j}A_{j},
\end{equation}%
\begin{equation}
\left( \partial ^{2}A_{k}-\partial _{k}\partial _{0}A_{0}+\partial
_{k}\partial _{i}A_{i}\right) =8{\cal B}D_{k}\left( D_{j}A_{j}+D_{0}A_{0}\right) .
\end{equation}
We choose the Coulomb gauge%
\begin{equation}
\partial _{r}A_{r}=0.
\end{equation}
The final equations are
\begin{equation}
\left( -\nabla ^{2}+8{\cal B}D_{0}^{2}\right) A_{0}=-8{\cal B}D_{0}D_{j}A_{j},
\end{equation}%
\begin{equation}
\left( \partial ^{2}A_{k}-\partial _{k}\partial _{0}A_{0}\right)
=8{\cal B}D_{k}\left( D_{j}A_{j}+D_{0}A_{0}\right) .
\end{equation}%
As in the usual electrodynamics $\;A_{0}$ can be expressed as an
instantaneous function of $\;A_{j}$
\begin{equation}
A_{0}=-8{\cal B}\frac{1}{\left( -\nabla ^{2}+8{\cal B}D_{0}^{2}\right) }D_{0}D_{j}A_{j}.
\end{equation}%
The final vector equation leads to
\begin{equation}
\partial ^{2}A_{k}=\left( -\nabla ^{2}\delta _{kl}+\partial _{k}\partial
_{l}\right) \frac{8{\cal B}}{\left( -\nabla ^{2}+8{\cal B}D_{0}^{2}\right) }%
D_{l}D_{j}A_{j},
\end{equation}%
where it is a direct matter to verify the Coulomb gauge condition.

The electromagnetic fields are given by
\begin{eqnarray}
\mathbf{A} &=&\{A^{i}\}, \\
\mathbf{E} &=&\{E_{i}\},\;\;E_{i}=f_{0i}=\partial _{0}A_{i}-\partial
_{i}A_{0},\;\;\mathbf{E=-}\partial _{0}\mathbf{A-\nabla }A^{0}, \\
\mathbf{B} &=&\{B_{i}\},\;\;B_{k}=-\frac{1}{2}\epsilon _{klm}f_{lm}=\epsilon
_{klm}\partial _{l}A^{m},\;\;\epsilon _{123}=+1,\;\;\mathbf{B=\nabla \times A},
\end{eqnarray}%
which reduces to
\begin{equation}
E_{r}=-i\omega A_{r}-ik_{r}A_{0},\;\;\;B_{k}=-i\epsilon _{klm}k_{l}A_{m}
\end{equation}%
in momentum space with the notation $\mathbf{k=}\left\{ k_{i}\right\} \;$in
the three dimensional subspace.

\subsection{Modified Maxwell equations}

They are
\begin{eqnarray}
\mathbf{\nabla \cdot E}+8{\cal B}\left( \mathbf{e\cdot \nabla }\right) (\ \mathbf{%
B\cdot b}-\mathbf{E\cdot e})=4\pi \rho &&, \\
\mathbf{\nabla \times }(\mathbf{B}+8{\cal B}\mathbf{b}(\ \mathbf{B\cdot b}-\mathbf{%
E\cdot e}))-\frac{1}{c}\frac{\partial }{\partial t}[\mathbf{E}+8{\cal B}\mathbf{e}%
(\ \mathbf{B\cdot b}-\mathbf{E\cdot e})]=\frac{4\pi }{c}\mathbf{J} &&, \\
\mathbf{\nabla \cdot B}=0 &&, \\
\mathbf{\nabla \times E}+\frac{1}{c}\frac{\partial \mathbf{B}}{\partial t}=0
&&.
\end{eqnarray}%
In the notation of electrodynamics in a medium we can introduce
\begin{equation}
\mathbf{D}=\mathbf{E+}8{\cal B}\mathbf{e}(\ \mathbf{B\cdot b}-\mathbf{E\cdot e}%
),\;\;\;\;\mathbf{H}=\mathbf{B}+8{\cal B}\mathbf{b}(\mathbf{B\cdot b}-\mathbf{%
E\cdot e}),
\end{equation}%
such that
\begin{eqnarray}
\mathbf{\nabla \cdot D} &=&4\pi \rho ,\;\;\;\;\mathbf{\nabla \times H-}\frac{%
1}{c}\frac{\partial \mathbf{D}}{\partial t}\mathbf{=}\frac{4\pi }{c}\mathbf{%
J,} \\
\mathbf{\nabla \cdot B} &\mathbf{=}&0,\;\;\;\;\mathbf{\nabla \times E+}\frac{%
1}{c}\frac{\partial \mathbf{B}}{\partial t}\mathbf{=}0.
\end{eqnarray}%
In components
\begin{eqnarray}
D_{i} &=&(\delta _{ij}-8{\cal B}e_{i}e_{j})E_{j}+8{\cal B}e_{i}b_{j}B_{j}, \\
H_{i} &=&(\delta _{ij}+8{\cal B}b_{i}b_{j})B_{j}-8{\cal B}b_{i}e_{j}E_{j}.
\end{eqnarray}%
Particular cases are
\begin{eqnarray}
b_{j} &=&0\rightarrow D_{i}=(\delta _{ij}-8{\cal B}e_{i}e_{j})E_{j},\;H_{i}=B_{i}\;,
\\
e_{i} &=&0\rightarrow \;\;D_{i}=E_{i},\;\;\;H_{i}=(\delta
_{ij}+8{\cal B}b_{i}b_{j})B_{j}.
\end{eqnarray}%
The first case induces a modification in Coulomb law$\;$
\begin{equation}
\mathbf{B=0,\;\;\;\nabla \cdot }\left( \mathbf{E-}8{\cal B}\mathbf{e}\left( \mathbf{%
e\cdot E}\right) \right) =4\pi \rho ,\;\;\mathbf{E}=-\mathbf{\nabla }\Phi ,
\end{equation}%
\begin{equation}
\left( \nabla ^{2}-8{\cal B}\left( e_{i}\partial _{i}\right) ^{2}\right) \Phi
=-4\pi \rho.
\end{equation}
\section{Summary of the dispersion relations and electromagnetic fields}
Here we use the previous decomposition of the background field $D_{\mu \nu }$
in terms of three dimensional components together with the $3+1$
formulation described in subsection (IV-B), to search for plane wave
solutions of the Maxwell equations. We assume the space-time dependence of
any field to be proportional to $e^{i(\mathbf{k\cdot x}-\omega t)}$, where $%
\mathbf{k}$ is the momentum of the wave and we choose the
potential $\mathbf{A}$ in the  Coulomb gauge. The notation is
\begin{equation}
\mathbf{b=\{}b_{m}\mathbf{\},\;\;e\mathbf{=\{}}e_{m}\mathbf{\mathbf{\}}\;.}
\end{equation}%
The expressions of Eqs. (\ref{P1}) , (\ref{P2})\ in momentum\ space are
\begin{eqnarray}
\left\{ D_{k}\right\} &=&\mathbf{D}=-i\left[ \omega \mathbf{e}-\mathbf{k}%
\times \mathbf{b}\right],  \label{P11} \\
D_{0} &=&-i\mathbf{e\cdot k}.  \label{P22}
\end{eqnarray}
There are two main cases

(i) $D_{j}A_{j}=0$: In this case the dispersion relation is $\omega =|%
\mathbf{k}|$. The fields are
\begin{eqnarray}
\mathbf{B}=\gamma \{(\mathbf{e}\cdot \mathbf{\hat{k}})\mathbf{\hat{k}}-%
\mathbf{e}+\mathbf{b}\times \mathbf{\hat{k}}\} &&,  \label{EDA0} \\
\mathbf{E}=\gamma \{(\mathbf{b}\cdot \mathbf{\hat{k}})\mathbf{\hat{k}}-%
\mathbf{b}-\mathbf{e}\times \mathbf{\hat{k}}\} &&.
\end{eqnarray}

(ii) $D_{j}A_{j}\neq 0$ : In this case the dispersion relation and the
fields are%
\begin{equation}
\omega =|\mathbf{k|}\frac{\sqrt{256{\cal B}^{2}(\mathbf{\hat{k}}\cdot \left(
\mathbf{b}\times \mathbf{e}\right) )^{2}-4(1-8{\cal B}\mathbf{e}^{2})\{8{\cal B}[(\mathbf{%
\hat{k}}\cdot \mathbf{e})^{2}-(\mathbf{\hat{k}}\times \mathbf{b})^{2}]-1\}}%
-16{\cal B}\mathbf{\hat{k}\cdot }\left( \mathbf{b}\times \mathbf{e}\right) }{2(1-8{\cal B}%
\mathbf{e}^{2})},  \label{GENDR}
\end{equation}%
\begin{eqnarray}
\mathbf{B}=\gamma \{\mathbf{k}\times (\mathbf{k}\times \mathbf{b})-\omega
\mathbf{k}\times \mathbf{e}\} &&, \\
\mathbf{E}=\gamma \{\omega \mathbf{k}\times \mathbf{b}-\omega ^{2}\mathbf{e}+%
\mathbf{k}(\mathbf{e}\cdot \mathbf{k})\} &&.
\end{eqnarray}%
In both cases $\gamma $ is an arbitrary constant. For small ${\cal B}$, the
dispersion relation (\ref{GENDR}) reduces to
\begin{equation}
\omega =|\mathbf{k}|\left[ 1+{\cal B}\left( 8\mathbf{\hat{k}}\cdot \left( \mathbf{e}%
\times \mathbf{b}\right) +4(\mathbf{\hat{k}}\times \mathbf{b})^{2}+4(\mathbf{%
\hat{k}}\times \mathbf{e})^{2}\right) \right] \,.  \label{SMALLBDR}
\end{equation}
The above expression clearly shows that the model is stable
in the small Lorentz invariance violation approximation, where the quantities $ {\cal B}e^2, {\cal B}b^2,  {\cal B}|\mathbf{e}||\mathbf{b}|$ are very small compared to one.

The anisotropic velocity of light arising from the above dispersion relation
is
\begin{eqnarray}
\mathbf{\nabla }_{\mathbf{k}}\omega &=&\mathbf{c(\hat{k})}=\mathbf{\hat{k}}%
\left( 1+8{\cal B}\left( b^{2}+e^{2}\right) -4{\cal B}\left( (\mathbf{\hat{k}}\times
\mathbf{b})^{2}+(\mathbf{\hat{k}}\times \mathbf{e})^{2}\right) \right)
\nonumber \\
&&+8{\cal B}\left( \mathbf{e}\times \mathbf{b}\right) \;\mathbf{-}8{\cal B}\left( \mathbf{%
b\cdot \hat{k}}\right) \mathbf{b-}8{\cal B}\left( \mathbf{e\cdot \hat{k}}\right)
\mathbf{e}\,.  \label{CANISOTROPIC}
\end{eqnarray}
Next we consider the separate cases where $\mathbf{e}\neq 0,\;\mathbf{b}=0$,
$\;\mathbf{e}=0,\;\mathbf{b}\neq 0,\;\;$and $\mathbf{e}\neq 0,\;\mathbf{b}%
\neq 0$. The results are:
\subsection{Case I: $\mathbf{b}=0,\;\;\mathbf{e}\neq 0$}
\subsubsection{CASE\ (I-1)\ \ $\mathbf{e\cdot A}$\ $\neq 0,\;\;\mathbf{%
D\cdot A}\ \neq 0\;$}
We find the dispersion relation
\begin{equation}
\omega _{1}^{2}=\frac{1-8{\cal B}\mathbf{e}^{2}\cos ^{2}\theta }{1-8{\cal B}\mathbf{e}^{2}}%
\mathbf{k}^{2},  \label{MDRE}
\end{equation}%
which agrees with the one obtained from (\ref{DR1}). Here $\theta$ is the
angle between $\mathbf{k}$ and $\mathbf{e}$. Assuming the corrections
terms to be very small we can rewrite (\ref{MDRE}) as
\begin{equation}
\omega _{1}=|\mathbf{k|}\left( 1+4{\cal B}\mathbf{e}^{2}\sin ^{2}\theta \right).
\label{MDRE1}
\end{equation}%
The electromagnetic fields in momentum space are
\begin{equation}
\mathbf{E}=\alpha \left[ \mathbf{e}-\frac{\left( \mathbf{e\cdot k}\right) }{%
\omega _{1}^{2}}\mathbf{k}\right] ,\qquad \mathbf{B}=\frac{\alpha }{\omega _{1}}%
\left[ \mathbf{k\times e}\right] ,
\end{equation}%
where\ $\alpha$ is a dimensionless constant. The following orthogonality
relations are satisfied
\begin{eqnarray}
\mathbf{E\cdot B} =0, \quad \mathbf{\hat{k}\cdot E} \neq 0,\quad \mathbf{\hat{k}\cdot B}=0,
\end{eqnarray}%
together with the properties%
\begin{equation}
\mathbf{E\times B\sim \;}\left[ \mathbf{\mathbf{k}-}8{\cal B}\mathbf{\left( \mathbf{%
e\cdot k}\right) e}\right] \equiv \mathbf{m,}
\end{equation}%
\begin{equation}
\mathbf{m\cdot B}=0,\;\;\mathbf{m\cdot E}=0.
\end{equation}%
The ratio between the amplitudes is
\begin{equation}
\frac{|\mathbf{E|}^{2}}{|\mathbf{B|}^{2}}=\frac{\sin ^{2}\theta +\left( 1-8{\cal B}%
\mathbf{e}^{2}\right) ^{2}\cos ^{2}\theta }{\left( 1-8{\cal B}\mathbf{e}^{2}\cos
^{2}\theta \right) \left( 1-8{\cal B}\mathbf{e}^{2}\right) }.  \label{RAMPE}
\end{equation}
\subsubsection{CASE\ (I-2)$\;\;\mathbf{e\cdot A}$\ $=0,\;\;\mathbf{D\cdot A}$%
\ $=0$}
The \ corresponding dispersion relation is
\begin{equation}
\omega _{2}^{2}=\mathbf{k}^{2},
\end{equation}%
in accordance with (\ref{DR2}), with the electromagnetic fields
\begin{equation}
\mathbf{E}=\beta \left( \mathbf{\hat{k}}\times \mathbf{e}\right) ,\;\;\;%
\mathbf{B}=\beta \;\mathbf{\hat{k}}\times \left( \mathbf{\hat{k}}\times
\mathbf{e}\right) ,
\end{equation}%
where $\beta $ \ is an arbitrary dimensionless constant. Also we have%
\begin{equation}
|\mathbf{E|=|B|=}\beta \mathbf{|e|}\sin \theta .
\end{equation}

The orthogonality relations are
\begin{equation}
\mathbf{\hat{k}\cdot E=}0,\quad \mathbf{\hat{k}\cdot B=}0,\quad \mathbf{E\cdot B=0},
\end{equation}%
similarly to the standard case.\ Also
\begin{equation}
\frac{|\mathbf{E|}^{2}}{|\mathbf{B|}^{2}}=1.
\end{equation}

\subsection{CASE II: \ $\mathbf{e}=0,\;\;\mathbf{b}\neq 0,\;\;D_{0}=0,\;$}
\subsubsection{CASE\ (II-1) $\mathbf{b\cdot A}$\ $\neq 0\;\Longrightarrow \;%
\mathbf{D\cdot A}=0$}
We find the dispersion relation
\begin{equation}
\omega _{3}^{2}=\mathbf{k}^{2}.
\end{equation}%
and we have%
\begin{equation}
\mathbf{E}=\gamma \left[ \mathbf{\hat{k}\times }\left( \mathbf{\hat{k}\times
b}\right) \right], \quad  \mathbf{B}=-\gamma \left( \mathbf{\hat{k}\times b},
\right)
\end{equation}
with $\gamma $ an arbitrary dimensionless constant. Also
\begin{equation}
\frac{|\mathbf{E|}^{2}}{|\mathbf{B|}^{2}}=1, \quad \mathbf{\hat{k}\cdot E = \hat{k}\cdot B}=0,\quad \mathbf{E\cdot B}=0.
\end{equation}%
\subsubsection{CASE\ (II-2)$\;\mathbf{b}\cdot \mathbf{A}=0\;\Longrightarrow
\mathbf{D\cdot A\neq 0}$}
Here the dispersion relation is
\begin{equation}
\omega _{4}^{2}=\mathbf{k}^{2}\left( 1+8{\cal B}\mathbf{b}^{2}\sin ^{2}\theta
\right) ,  \label{MDRE2}
\end{equation}%
with $\theta$ being the angle between $\mathbf{k}$ and $\mathbf{b}$. The above
result agrees with that obtained from (\ref{DR1}). \ The electromagnetic
fields are determined, up to the arbitrary constant $\delta $,  by
\begin{eqnarray}
\mathbf{E} &=&\frac{\delta }{\omega _{4}}\left( \mathbf{k\times b}\right) ,
\\
\mathbf{B} &=&\frac{\delta }{\omega _{4}^{2}}\;\mathbf{k}\times \left(
\mathbf{k\times b}\right)
\end{eqnarray}
and they satisfy
\begin{equation}
\mathbf{\hat{k}}\cdot \mathbf{E}=0, \quad \mathbf{\hat{k}} \cdot \mathbf{B}=0,\quad
\mathbf{E} \cdot \mathbf{B}=0,
\end{equation}%
\begin{equation}
\frac{|\mathbf{E|}^{2}}{|\mathbf{B|}^{2}}=\frac{\omega _{4}^{2}}{|\mathbf{k|}%
^{2}}=\left( 1+8{\cal B}\mathbf{b}^{2}\sin ^{2}\theta \right) .
\end{equation}
Notice that half of the cases previously considered violate rotational
invariance, with the exception of those having unmodified dispersion
relations together with standard properties of the propagating
electromagnetic fields.
\subsection{CASE\ III:\ $\;\mathbf{b}^{2}=\mathbf{e}^{2},\;\;\mathbf{e\cdot }\;%
\mathbf{b=0}$}
We consider this case corresponding to a plane wave vacuum because
the final theory remains invariant under the three-parameter subgroup HOM(2).

We choose a reference frame such that
\begin{equation}
\mathbf{\hat{n}}\mathbf{=}(1,0,0),\;\;\mathbf{b}\mathbf{=}(0,0,b),\;\;\;%
\mathbf{e}\mathbf{=}(0,sb,0),\;s=\pm 1,
\end{equation}%
\begin{equation}
\mathbf{k\cdot A=0}
\end{equation}%
\begin{equation}
D_{j}A_{j}=ib\left( \left( k_{1}-\omega s\right) A_{2}-k_{2}A_{1}\right)
\end{equation}
\subsubsection{Case $D_{k}A_{k}=0.$}
The dispersion relations are the usual ones. From the master formulae
(116-117), the fields are
\begin{eqnarray}
\mathbf{B}=\gamma \{(s(\mathbf{b\times \hat{n}})\cdot \mathbf{\hat{k}})%
\mathbf{\hat{k}}-s(\mathbf{b}\times \mathbf{\hat{n}})+\mathbf{b}\times
\mathbf{\hat{k}}\} &&, \\
\mathbf{E}=\gamma \{(\mathbf{b}\cdot \mathbf{\hat{k}})\mathbf{\hat{k}}-%
\mathbf{b}+s\mathbf{\hat{k}}\times (\mathbf{b}\times \mathbf{\hat{n}})\} &&.
\end{eqnarray}
\subsubsection{Case $D_{k}A_{k}\neq 0.$}
The exact modified dispersion relations are
\begin{equation}
\omega _{5}=\frac{1}{\left( 1-8{\cal B}b^{2}\right) }\left( \sqrt{\left( \mathbf{k}%
^{2}-8{\cal B}\mathbf{b}^{2}\left( \mathbf{k}^{2}-k_{1}^{2}\right) \right) }%
+8{\cal B}b^{2}sk_{1}\right).
\label{FULLMDR}
\end{equation}%
In the small violation approximation Eq. (\ref{FULLMDR})  reduces to
\begin{equation}
\omega _{5}=|\mathbf{k}|\left( 1+4{\cal B}b^{2}\left[ 1+s\frac{k_{1}}{|\mathbf{k}|}%
\right] ^{2}\right) .\;\;\;
\end{equation}

From the formulae (118-119), the fields are
\begin{eqnarray}
\mathbf{B}=\gamma \{\mathbf{\hat{k}}\times (\mathbf{\hat{k}}\times \mathbf{b}%
)-\omega _{5}s\mathbf{\hat{k}}\times (\mathbf{b}\times \mathbf{\hat{n}})\}
&&, \\
\mathbf{E}=\gamma \{\omega _{5}\left( \mathbf{\hat{k}}\times \mathbf{b}%
\right) -\omega _{5}^{2}s(\mathbf{b}\times \mathbf{\hat{n}})+s\mathbf{\hat{k}%
}((\mathbf{b}\times \mathbf{\hat{n}})\cdot \mathbf{\hat{k}})\} &&.
\end{eqnarray}

\section{THE MODEL AS A SECTOR OF THE SME}

In order to impose some bounds upon the parameters of the model it is
convenient to recast the quadratic sector of the action (\ref{LAGFIN1}) in
the language of the Standard Model Extension and to make use of the numerous
experimental constraints derived from it ( see for example Ref. \cite{KOSTELECKYED}) and
summarized in Table II:\ Photon-sector summary, of \ Ref. \cite{TKOSTELECKY}.
To begin with, let us recall the dimension of the fields and parameters
involved in the model
\begin{equation}
\left[ A\right] =m,\hspace{0.75em}\hspace{0.75em}\left[ e\right] =\left[ b%
\right] =m^{2},\hspace{0.75em}\hspace{0.75em}\left[ {\cal B}\right] =\frac{1}{m^{4}}%
,\hspace{0.75em}\hspace{0.75em}\left[ {\cal B} e^{2}\right] =0.
\end{equation}%
Next we make the identification
\begin{equation}
-{\cal B}\left( f_{\mu \nu }D^{\mu \nu }\right) ^{2}=-\frac{1}{4}\left(
k_{F}\right) ^{\kappa \lambda \mu \nu }f_{\kappa \lambda }f_{\mu \nu },
\end{equation}%
where the tensor $\left( k_{F}\right) ^{\kappa \lambda \mu \nu }$, with 19
independent components, has all the symmetries of the Riemann tensor and a
vanishing double trace. In terms of the matrix elements $D_{\mu \nu }\;$%
characterizing the vacuum expectation values of the electromagnetic tensor
we obtain
\begin{equation}
\left( k_{F}\right) ^{\kappa \lambda \mu \nu }=4{\cal B}\left[ D^{\kappa \lambda
}D^{\mu \nu }+\frac{1}{2}\left( D^{\kappa \mu }D^{\lambda \nu }-D^{\lambda
\mu }D^{\kappa \nu }\right) \right] -\frac{1}{2}{\cal B}D^{2}\left( \eta ^{\kappa
\mu }\eta ^{\lambda \nu }-\eta ^{\lambda \mu }\eta ^{\kappa \nu }\right) ,
\end{equation}%
where%
\begin{equation}
D^{2}\equiv D_{\alpha \beta }D^{\alpha \beta }=2(\mathbf{b}^{2}-\mathbf{e}%
^{2}).
\end{equation}

Next we have to identify the appropriate combinations of the components of $%
\left( k_{F}\right) ^{\kappa \lambda \mu \nu }\;$which are bounded. A first
step in that direction constitutes the definitions of the components $\left(
\kappa _{DE}\right) ^{jk},\;\left( \kappa _{HB}\right) ^{jk}\;$and $\left(
\kappa _{DB}\right) ^{jk}=-\left( \kappa _{HE}\right) ^{kj}$, which are
written in terms of our parameters $\mathbf{e},\mathbf{b}\;$and ${\cal B}$\ as
follows
\begin{equation}
\left( \kappa _{DE}\right) ^{jk}\equiv -2\left( k_{F}\right)
^{0j0k}=-12{\cal B}e_{j}e_{k}-{\cal B}D^{2}\delta _{jk},
\end{equation}%
\begin{eqnarray}
\left( \kappa _{HB}\right) ^{jk} &\equiv &\frac{1}{2}\epsilon _{jpq}\epsilon
_{krs}\left( k_{F}\right) ^{pqrs}=12{\cal B}b_{j}b_{k}-{\cal B}D^{2}\delta _{jk}, \\
\left( \kappa _{DB}\right) ^{jk} &=&-\left( \kappa _{HE}\right)
^{kj}=\epsilon _{kpq}(k_{F})^{0jpq}=12{\cal B}(e_{j}b_{k}-\frac{1}{3}\left( \mathbf{%
e\cdot b}\right) \delta _{jk}),\;\;\;\;tr\left[ \kappa _{DB}^{jk}\right] =0. \nonumber
\\
\end{eqnarray}%
The final combinations in terms of which the bounds are presented turn out to be
\begin{eqnarray}
\left( \bar{\kappa}_{e+}\right) ^{jk} &=&\frac{1}{2}\left( \kappa
_{DE}+\kappa _{HB}\right) ^{jk}=6{\cal B}\, \left[
b_{j}b_{k}-e_{j}e_{k}-\frac{1}{3}\left( \mathbf{b}^{2}-\mathbf{e}^{2}\right)
\delta _{jk}\right] \;<10^{-32},  \label{DEF1} \\
\left( \bar{\kappa}_{e-}\right) ^{jk} &=&\frac{1}{2}\left( \kappa
_{DE}-\kappa _{HB}\right) ^{jk}-\frac{1}{3}\delta _{jk}tr(\kappa _{DE})
\nonumber \\
&=&-6{\cal B}\left[ (b_{j}b_{k}+e_{j}e_{k})-\frac{1}{3}\delta _{jk}(\mathbf{b}^{2}+%
\mathbf{e}^{2})\right] \;<10^{-16}\;,  \label{DEF2} \\
(\bar{\kappa}_{o+})^{jk} &=&\frac{1}{2}\left( \kappa _{DB}+\kappa
_{HE}\right) ^{jk}=6{\cal B}(e_{j}b_{k}-e_{k}b_{j})\;<10^{-12},  \label{DEF3} \\
(\bar{\kappa}_{o-})^{jk} &=&\frac{1}{2}\left( \kappa _{DB}-\kappa
_{HE}\right) ^{jk}=6{\cal B}\left[ (e_{j}b_{k}+e_{k}b_{j})-\frac{2}{3}\left(
\mathbf{e\cdot b}\right) \delta _{jk}\right] \;<10^{-32},  \label{DEF4} \\
\bar{\kappa}_{tr} &=&\frac{1}{3}tr\left( k_{DE}\right) ^{j}=-2{\cal B}(\mathbf{e}%
^{2}+\mathbf{b}^{2})\;<10^{-15}.  \label{DEF5}
\end{eqnarray}%
The bounds for  $\bar{\kappa}_{tr}$ have recently been improved from $%
10^{-7}$ \cite{TKOSTELECKY}, to$\;10^{-11\;}$\cite{LEHNERT1}, and finally
to\ $10^{-15\;}$\cite{ALTSCHUL}. \ In quoting our results arising from the
bounds upon $\left( \bar{\kappa}_{e+}\right) ^{jk},\;(\bar{\kappa}%
_{o-})^{jk} $ and $\bar{\kappa}_{tr}$\ we have chosen our reference frame in
such a way that only the light-cone gets modified by the LIV parameters,
while the propagation of the fermioms is the standard one \cite{KOSTELECKYED}%
,\cite{FRAME}.

All matrices from (\ref{DEF1}) to (\ref{DEF4})\ are traceless, with $(\bar{%
\kappa}_{0+})^{jk}\;$being antisymmetric ($3$ independent components) while
the three remaining ones are symmetric ($5$ independent components each). \
The above combinations (\ref{DEF1}) to (\ref{DEF5})\ constitute a convenient
alternative way to display the $19$ independent components of the tensor $%
\left( k_{F}\right) ^{\alpha \beta \mu \nu }.\;\;$The bounds are obtained in
terms of such combinations referred to the Standard Inertial Reference Frame
defined in Ref. \cite{KOSTELECKYED} and are written at the end of each of
the above equations.

Notice that the bilinears which are odd under the duality transformations
\begin{equation}
{\mathbf{e}\rightarrow \mathbf{b},\;\;\;\;\mathbf{b}\rightarrow \mathbf{-e}},
\end{equation}%
are much more suppressed than those which are even. In this way, even
though our model is not duality invariant, this transformation seems to
explain the above mentioned hierarchy in the LIV parameters exhibited in
Eqs. (\ref{DEF1})-(\ref{DEF5}).
\begin{figure}[tbp]
\includegraphics[height=8cm]{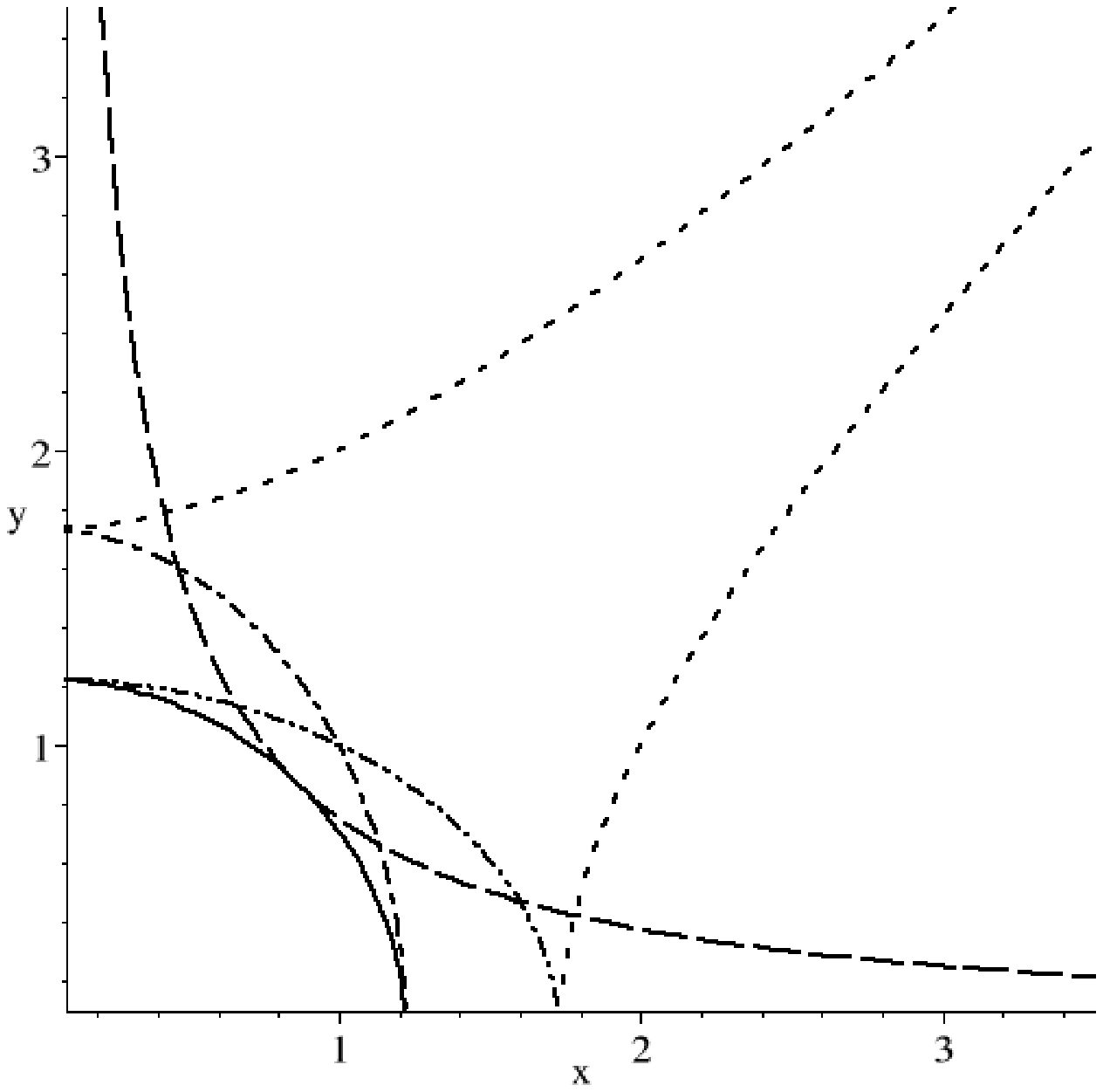}
\caption{ Boundaries of the allowed region obtained from the constraints in
the parameters ${\tilde{\protect\kappa}}_{e+}^{jk}$ and ${\tilde{\protect%
\kappa}}_{o-}^{jk}$. The allowed region is on the inside the dashed,
dot-dashed and dot-dot-dashed lines. An excellent approximation for it is
the sector of the circle shown in solid line \cite{RLFIG}.}
\end{figure}
In order to obtain some specific consequences of the bounds (\ref%
{DEF1}) to (\ref{DEF5}) it is convenient to express them in the coordinate
system defined by (\ref{PREFCOORD}). Also we introduce the notation%
\begin{equation}
\mathbf{x}=\sqrt{6{\cal B}}\,\mathbf{e\times }10^{16},\;\;\;\mathbf{y}=\sqrt{6{\cal B}}%
\, \mathbf{b\times }10^{16},\;\;\quad x=|\mathbf{x}|,\quad y=|\mathbf{y}|
\end{equation}%
We consider only absolute values of the related quantities and we focus upon
the stringent  constraints  in  Eqs. (\ref{DEF1}) and (\ref{DEF4}). The
non-trivial contributions are
\begin{eqnarray}
\left| \left( \tilde{\kappa}_{e+}\right) ^{11}\right| &=&\left|
(x^{2}-y^{2})\right| <3,  \label{KE+1} \\
\left| \left( \tilde{\kappa}_{e+}\right) ^{22}\right| &=&\left| \left(
1-3\sin ^{2}\psi \right) x^{2}-y^{2}\right| < 3,  \label{KE+2} \\
\left| \left( \tilde{\kappa}_{e+}\right) ^{33}\right| &=&\left|
2y^{2}+\left( 1-3\cos ^{2}\psi \right) x^{2}\right| < 3,  \label{KE+3} \\
\left| \left( \tilde{\kappa}_{e+}\right) ^{23}\right| &=&\left| x^{2}\sin
2\psi \right| < 2,  \label{KE+4}
\end{eqnarray}%
together with%
\begin{eqnarray}
\left( \tilde{\kappa}_{o-}\right) ^{11} &=&\left( \tilde{\kappa}_{o-}\right)
^{22}=\left| xy\cos \psi \right| <\frac{3}{2},  \label{KO-1} \\
\left( \tilde{\kappa}_{o-}\right) ^{33} &=&\left| xy\cos \psi \right| <\frac{%
3}{4},  \label{KO-2} \\
\left( \tilde{\kappa}_{o-}\right) ^{23} &=&\left| xy\sin \psi \right| < 1.
\label{KO-3}
\end{eqnarray}%
The allowed region in the $(x-y)$\ plane is shown in Fig.1, where we plot
the boundaries of the corresponding inequalities. In expressions (\ref%
{KE+4}), (\ref{KO-1})-(\ref{KO-3}) we consider the lower bound corresponding
to the maximum value of the trigonometric function on the LHS of the
inequality. This leads to
\begin{equation}
x<\sqrt{2},\;\;\left| xy\right| <\frac{3}{4}<1<\frac{3}{2}.
\end{equation}%
The boundary\ $xy=3/4$ is plotted in dashed line. The most
stringent bound from Eq. (\ref{KE+3}) results when $\psi =\pi /2$ and corresponds to
\begin{equation}
y<\sqrt{\frac{3-x^{2}}{2}}.
\end{equation}%
The boundary is  shown in dot-dot-dashed line. The most stringent
bound from (\ref{KE+2}) arises again from $\psi =\pi /2$  and corresponds to
\begin{equation}
y<\sqrt{3-2x^{2}},
\end{equation}%
which boundary is plotted in dot-dashed line. The expression (\ref{KE+1}) does not
provide additional bounds and is plotted for completeness. The corresponding
boundaries are $y_{\pm }=\sqrt{x^{2}\pm 3}$  shown in dotted lines in Fig.
1.\ An upper bound including all previous ones is given by the interior of
the circle
\begin{equation}
y=\sqrt{\frac{3}{2}-x^{2}},
\end{equation}%
shown in solid line in Fig. 1  and which translates into  the restriction
\begin{equation}
{\cal B}(\mathbf{e}^{2}+\mathbf{b}^{2})=\frac{1}{M_{{\cal B}}^{4}}(\mathbf{e}^{2}+\mathbf{b%
}^{2})\;<2.5\times 10^{-33}.  \label{E2pB2}
\end{equation}
This bound incorporates all the constraints  established in Eqs.(\ref{DEF1})-(\ref{DEF5}).\ All the
above relations are valid in the Standard Inertial Reference Frame\ centered
in the Sun \cite{KOSTELECKYED}.

\section{SUMMARY AND FINAL COMMENTS}

We  study a novel way of implementing a model with spontaneously
broken Lorentz symmetry by introducing a constant vacuum expectation value
(VEV) of the field strength $\langle F_{\mu \nu }\rangle =C_{\mu \nu }$ %
instead of imposing a VEV of the electromagnetic potential $\langle A_{\mu
}\rangle$ as frequently investigated in the literature. In this way, our
model preserves gauge invariance from the very beginning. We start from
the effective Lagrangian (\ref{LAGINI}) describing non-linear electrodynamics and
containing a potential $V_{eff}(F_{\mu \nu })$, which is  argued to arise after integrating massive
gauge bosons and fermions in an underlying conventional theory. This approach is an extension of the model in
Refs. \cite{Bjorken,KT}, which includes all quartic fermion interactions allowed by the Dirac algebra.
For simplicity we have considered only one fermion species. The subsequent integration of them produces
the  effective gauge invariant potential which is bounded in the high-intensity field limit and has an stable minimum.
In order to investigate the dynamical consequences of the electrodynamics constructed around such minimum we have chosen  the standard Ginzburg-Landau  parametrization for   $V_{eff}(F_{\mu \nu })$.
We have verified that the  extremum
conditions in the potential corresponds to the extremum conditions of the
energy  arising from (\ref{LAGINI}), under the requirement that the involved
fields are constant at the extreme points. Next we expand the fields around
the minimum and arrive to the spontaneously broken action (\ref{LAG2}) from
where we can further discuss the resulting non-linear electrodynamics,
which differs from the standard one described in Ref. \cite{PLEBANSKI},
for example. The action (\ref{LAG2})\ contains the field strength $a_{\mu
\nu }\;(\mathbf{E}\ ,\;\mathbf{B})$, together with the auxiliary field $\bar{%
X}_{\mu }$ which equations of motion demand that the two-form $a$ is such
that $da=0$, thus introducing the electromagnetic potentials. The
constitutive relations expressing the electromagnetic excitations $\mathbf{D}
$ and $\mathbf{H}$ , are identified via the equations of motions (\ref%
{EQMOT1}),\ (\ref{EQMOT3}) and provide a direct connection with the field
strengths $\mathbf{E}$  and $\mathbf{B}$. After eliminating the auxiliary
field $\bar{X}_{\mu }$ and writing the most general solution to the
condition $da=0$, we arrive at the action (\ref{LAGFIN1}),\ which is the
starting point of our subsequent analysis. Some redefinitions allow the
symmetry breaking parameters to be characterized by the constant antisymmetrical
tensor $D_{\mu \nu }$ (proportional  to the VEV $C_{\mu \nu }$), plus an additional constant ${\cal B}$.

The search for the subgroups under which the model remains invariant after
the symmetry breaking  includes the generator of dilatations plus the
standard Lorentz generators. In this restricted subalgebra the former
commutes with the remaining ones, which allows its realization as a multiple
of the identity. There is only one case, given in the subsubsection IV.E-2,
which incorporates this generator, resulting in the breaking of the Lorentz
group to the three generators subgroup $HOM(2)$. All the other cases break
to a subgroup isomorphic to $T(2)$, with two generators.

The modified photon dispersion relations together with the corresponding
plane wave polarizations are classified according to the value of $
D_{i\alpha }k^{\alpha}A^{i}$, where $k^{\alpha }$ is the plane wave four
momentum and $A^{i}$ is the corresponding vector potential in the Coulomb
gauge. The amplitudes of the propagating fields are written in terms of those
parameterizing  $D_{\mu \nu }$. The case $D_{i\alpha }k^{\alpha
}A^{i}=0$ leads to dispersion relations $\omega =|\mathbf{k}|\;$with the
triad\ $\mathbf{E},$ $\mathbf{B,\;k\;}$having standard orthogonality
properties. The situation $D_{i\alpha }k^{\alpha }A^{i}\neq 0\;$produces\
the following properties for the vectors involved: $\mathbf{E}$ is
perpendicular to $\mathbf{B}$, $\mathbf{B}$\textbf{\ }is perpendicular to $%
\mathbf{k}$, \ but $\mathbf{E}$ is not necessarily orthogonal to $\mathbf{k}$%
. The corresponding dispersion relations are\ of the form%
\begin{equation}
\omega _{\pm }=|\mathbf{k}|\;F_{\pm },
\end{equation}%
where $F_{\pm }$ is independent of $|\mathbf{k}|,$ being only function of
the angles between\ $\mathbf{k}$ and $\mathbf{e}$ ,\ $\mathbf{k}$ and $%
\mathbf{b,\;}$ $\mathbf{k\;}$and $\mathbf{e\times b}$.\ \ In this way our
model predicts anisotropy in the speed of light. Also, as shown in Eq. (\ref{SMALLBDR}) the model is stable
in the small Lorentz invariance violation approximation where the quantities $ {\cal B}e^2, {\cal B}b^2,  {\cal B}|\mathbf{e}||\mathbf{b}|$ are very small compared to one. This is validated by the bound ${\cal B}(e^2+b^2) < 2.5 \times 10^{-33}$ derived from Fig.1.

In order to make a more quantitative statement about this anisotropy let us
consider the different possibilities arising from the velocity (\ref%
{CANISOTROPIC}). To the leading order ${\cal B}|\mathbf{e}|^{2}$, ${\cal B}|\mathbf{b}|^{2}$, ${\cal B}|\mathbf{e}||\mathbf{b}|$ the
magnitude of such velocity is
\begin{equation}
c(\mathbf{\hat{k}})=1+8{\cal B}\left( e^{2}+b^{2}\right) -4{\cal B}\left( \left( \mathbf{%
b\cdot \hat{k}}\right) ^{2}+\left( \mathbf{e\cdot \hat{k}}\right) ^{2}-2%
\mathbf{\hat{k}\cdot }\left( \mathbf{e}\times \mathbf{b}\right) \right),
\label{ONEWAYC}
\end{equation}%
which we take as the one-way light speed in the direction$\;\mathbf{\hat{k},}
$ predicted by our model. The corresponding two-way light speed is then%
\begin{equation}
c_{TW}(\mathbf{\hat{k}})=\frac{1}{2}\left( c(\mathbf{\hat{k}})+c(-\mathbf{%
\hat{k}})\right) =1+8{\cal B}\left( e^{2}+b^{2}\right) -4{\cal B}\left( \left( \mathbf{%
b\cdot \hat{k}}\right) ^{2}+\left( \mathbf{e\cdot \hat{k}}\right) ^{2}\right).
\label{TWOWAYC}
\end{equation}%
A standard measure for the anisotropy of the speed of light is the
comparison of the two-way velocities in perpendicular directions. We single
out the vector $\left( \mathbf{\hat{e}}\times \mathbf{\hat{b}}\right) =%
\mathbf{\hat{n}\;}$in our reference frame, which plays an analogous role to
the relative velocity among the ether frame and the laboratory frame in the
Robertson-Mansouri-Sexl (RMS) type of analysis \cite{ROBERTSON}, \cite{MS}.
The vectors $\mathbf{\hat{k}\;}$and $\mathbf{\hat{n}}$ form a plane in which
we define the vector $\mathbf{\hat{q}\;}$perpendicular to $\mathbf{\hat{k}\;}
$given by
\begin{equation}
\mathbf{\hat{q}=\hat{k}\times }\left( \mathbf{\hat{k}\times \hat{n}}\right)
=\left( \mathbf{\hat{k}\cdot \hat{n}}\right) \mathbf{\hat{k}-\hat{n}}.
\end{equation}%
Next we calculate the two-way light speed along this perpendicular direction
obtaining
\begin{equation}
c_{TW}(\mathbf{\hat{q}})=1+8{\cal B}\left( e^{2}+b^{2}\right) -4{\cal B}\left( \mathbf{%
\hat{k}\cdot \hat{n}}\right) ^{2}\left( b^{2}\left( \mathbf{\hat{b}}\cdot
\mathbf{\hat{k}}\right) ^{2}+e^{2}\left( \mathbf{\hat{e}}\cdot \mathbf{\hat{k%
}}\right) ^{2}\right).
\end{equation}%
An appropriate definition for the anisotropy of the speed of
light  in this model is
\begin{eqnarray}
\frac{\Delta c}{c} &\equiv &\left| c_{TW}(\mathbf{\hat{k}})-c_{TW}(\mathbf{%
\hat{q}})\right|, \\
\frac{\Delta c}{c} &=&\left| 4{\cal B}\left(1-\left(\mathbf{\hat{k}\cdot \hat{n}}%
\right) ^{2}\right)\left( b^{2}\left( \mathbf{\hat{b}\cdot \hat{k}}\right)
^{2}+e^{2}\left( \mathbf{\hat{e}\cdot \hat{k}}\right) ^{2}\right) \right| .
\end{eqnarray}%
From the last expression we obtain the bound
\begin{equation}
\frac{\Delta c}{c}<4{\cal B}\sin ^{2}\theta (\mathbf{b}^{2}+\mathbf{e}^{2})<4{\cal B}(%
\mathbf{b}^{2}+\mathbf{e}^{2})\;<10^{-32},  \label{BOUNDDC}
\end{equation}%
according to Eq.(\ref{E2pB2}), where $\theta$ is the angle between the vector
$\mathbf{\hat{n}}$ and $\mathbf{\hat{k}}$. The above anisotropy
measures the difference in the two-way speed of light propagating in
perpendicular directions in a given reference frame. The standard
Michelson-Morley type of analysis can be made by measuring first the time
difference along two perpendicular trajectories $L_{1}\;$(propagating with $%
\;c_{TW}(\mathbf{\hat{k}})$),\ and $L_{2}\;$(propagating with $\;c_{TW}(%
\mathbf{\hat{q}})$)%
\begin{equation}
\tau =\frac{L_{2}}{c_{TW}(\mathbf{\hat{q}})}-\frac{L_{1}}{c_{TW}(\mathbf{%
\hat{k}})}.
\end{equation}%
Subsequently a similar measurement is made by rotating the apparatus by $90$
degrees%
\begin{equation}
\tau ^{\prime }=\frac{L_{2}}{c_{TW}(-\mathbf{\hat{k}})}-\frac{L_{1}}{c_{TW}(%
\mathbf{\hat{q}})}.
\end{equation}%
Finally, the difference in optical paths
\begin{equation}
\Delta \tau =\left| \tau -\tau ^{\prime }\right| =4{\cal B}\left(
L_{2}+L_{1}\right) \sin ^{2}\theta \left( b^{2}\left( \mathbf{\hat{b}}\cdot
\mathbf{\hat{k}}\right) ^{2}+e^{2}\left( \mathbf{\hat{e}}\cdot \mathbf{\hat{k%
}}\right) ^{2}\right)
\end{equation}%
measures the change in the interference pattern of the Michelson-Morley
experiment, according to our model.

It is important to emphasize that the bound (\ref{BOUNDDC}) is not related
to the anisotropy in the speed of light induced by the passage from a
preferred frame (where propagation is isotropic) to another frame moving
with relative speed $\mathbf{v}$ and subjected to alternative methods of
time synchronization. As such, it cannot be directly compared with the RMS
type bounds appearing in the literature. The difference in two perpendicular
two-way speed of light in the RMS case is \cite{MS}
\begin{equation}
\frac{\delta c}{c}=\left| c_{TW}(\phi +\pi /2)-c_{TW}(\phi )\right| =\left(
\frac{v}{c}\right) ^{2}\left| (\beta -\frac{1}{2}+\delta )\cos 2\phi \right|
.
\end{equation}%
Here $\phi $ is the angle between the photon direction and the relative
speed $\mathbf{v}$. Recent bounds upon the Mansouri-Sexl parameter $(\beta -%
\frac{1}{2}+\delta )$ \cite{BOUNDSMS} are of the order of $10^{-10}$,
implying a RMS bound
\begin{equation}
\frac{\delta c}{c}\leq 10^{-16},
\end{equation}%
with the relative speed between the earth and the CMB reference frame,
where the cosmic background radiation looks isotropic, having the value $v=|\mathbf{v}%
|\simeq 300$ km/s.

Perhaps a more significant comparison of our results \ can be made with the
work of Ref. \cite{GRASIETAL}, where the Euler model of nonlinear
electrodynamics \cite{EULER} is used to predict the following  isotropic bound of the
one-way speed of light%
\begin{equation}
\frac{\tilde{\delta}c}{c}\leq 1.2\times 10^{-23}.
\end{equation}%
In our case this would correspond to averaging over all angles in (\ref%
{ONEWAYC}), leading to
\begin{equation}
\frac{\tilde{\delta}c}{c}\leq 8{\cal B}\left( e^{2}+b^{2}\right) \simeq 2\times
10^{-32}.
\end{equation}

Let us emphasize that in all our estimations we are assuming that the
involved references frames (CMB, Sun based, earth based) are concordant, in
such a way that  first order quantities in the Lorentz invariance
violation parameters remain of the same order in all of them.

Assuming that our background fields $\mathbf{e}$ and $\mathbf{b\;}$might
represent some galactic or intergalactic fields in the actual era, we obtain
a very reasonable bound for the magnetic intergalactic field by assuming
that the constant appearing in the action (\ref{LAGFIN1}) corresponds to an
energy density $\rho \;$%
\begin{equation}
\rho =\frac{1}{4}\left[ (1-D^{2}{\cal B}\right] D^{2}\simeq \frac{1}{2}\left(
\mathbf{b}^{2}-\mathbf{e}^{2}\right) ,
\end{equation}%
which we can associate with the cosmological constant, since it would
represent a global property of the universe. The fact that this constant is
positive favors $\left( \mathbf{b}^{2}-\mathbf{e}^{2}\right) \;>0$, so that
one can perform a passive Lorentz transformation to a reference frame where $%
\mathbf{e=0}$. Supposing further that this frame, which describes the
intergalactic fields, is concordant with the Standard Inertial Reference
Frame and taking the
upper limit \cite{CARROL}
\begin{equation}
|\rho _{\Lambda }|<\;10^{-48}\;\left( GeV\right) ^{4},
\end{equation}%
we obtain the bound
\begin{equation}
|\mathbf{b|\;<\;}5\times 10^{-5}\;Gauss,
\end{equation}%
which is consistent with observations of intergalactic magnetic fields. Let
us recall that $1\;Gauss=1.95\times 10^{-20}\;\left( GeV\right) ^{2}$. It is
amusing to observe that this bound on $|\mathbf{b}|$ translates into the
following condition for the mass scale $M_{B}$ introduced in Eq. (\ref{E2pB2}%
)
\begin{equation}
M_{\cal B}>1.4\times 10^{-4}\,\,GeV.
\end{equation}%
Recalling that the lightest charged particle is the electron with $%
m_{e}=5.1\times 10^{-4}\,\,GeV$, we notice that the above bound is
consistent with the QED Euler-Heisenberg type of non-linear corrections to
the photon interaction, which scale as $1/m^{4}$, with $m$ being the mass of
the charged particle contributing to the loop in the effective action \cite{GREINER}.

\section*{ The Appendix}

In this Appendix we extend the models in Refs \cite{Bjorken,KT} to include
all quartic gauge invariant fermionic interactions allowed by the Dirac algebra and look for
their contribution to the effective gauge invariant electromagnetic action by integrating the fermion field.
We heavily rely on the discussion and results of
Ref.\cite{DR}, which presents a complete and detailed description of
the standard electromagnetic calculation. We do not attempt to cite
the original papers, which references can be found in Ref.\cite{DR}.

After integrating the massive gauge bosons in an underlying conventional theory \cite{KT}, let us consider the effective Lagrangian \begin{equation}
L_{eff}=\bar{\Psi}\left( i\gamma ^{\mu }\left( \partial _{\mu
}+ieA_{\mu }\right) -m\right) \Psi -\sum_{M,a}\frac{r_{M}}{2\Lambda
^{2}}\left[ \left( \bar{\Psi}M_{a}\Psi \right) \left(
\bar{\Psi}M^{a}\Psi \right) \right] . \label{LINI}
\end{equation}%
The index $M: S, V, T, PV, PS$ labels the tensorial objects that constitute the
Dirac basis: Scalar, Vector, Tensor, Pseudovector and Pseudoscalar, respectively.
The generic index $a$ labels the covariant (contravariant) components of each  tensor
class. More details are given in Table I. We follow the conventions of
 Ref.\cite{BD}, with appropriate factors chosen in such a way that each current  $\left( \bar{%
\Psi}M^{a}\Psi \right)$ is real.

Our goal is to estimate the contributions
$W^{(2)}$ of the additional quartic
fermionic couplings to the standard effective electromagnetic
action $W_{EM}^{(1)}$. The latter is  obtained by integrating the quadratic contribution of the
fermions in Eq.(\ref{LINI}) and it
is given by
\begin{eqnarray}
\langle 0_{+}|0_{-}\rangle _{(1)}^{A} &=&\int \left[ D\bar{\Psi}\right] %
\left[ D\Psi \right] \;\exp \left[ i\int d^{4}x\;\bar{\Psi}\left(
i\gamma ^{\mu }\left( \partial _{\mu }+ieA_{\mu }\right) -m\right)
\Psi \right],
\nonumber \\
&=&\det (\bar{\Psi}\left( i\gamma ^{\mu }\left( \partial _{\mu
}+ieA_{\mu }\right) -m\right) \Psi )\equiv \exp \left[
iW_{EM}^{(1)}\right] =\exp \left[ i\int d^{4}x\;L_{EM}^{(1)}\right].
\end{eqnarray}%
To this end we introduce the gauge invariant auxiliary fields $(C_M)_a$,
via the couplings%
\begin{equation}
L_{eff}=\bar{\Psi}\left( i\gamma ^{\mu }\left( \partial _{\mu
}+ieA_{\mu
}\right) -m\right) \Psi +\sum_{M,a}\left[ \frac{p_{M}}{2}(C_M)_a(C_M)^{a}-q_{M}(C_M)_{a}%
\left( \bar{\Psi}M^{a}\Psi \right) \right] .
\end{equation}%
The equations of motion for the auxiliary fields are (we do not use the Einstein convention for repeated indices unless explicitly stated)
\begin{equation}
(C_M)_{a}=\frac{q_{M}}{p_{M}}\left( \bar{\Psi}M_{a}\Psi \right) ,
\end{equation}%
which reproduce (\ref{LINI})\ when
\begin{equation}
\frac{r_{M}}{\Lambda ^{2}}=\frac{q_{M}^{2}}{p_{M}}.
\end{equation}%
Next we integrate  over the auxiliary fields following the
standard steps which begin with the introduction of the currents
$(j_M)^{a}$ together with the replacements
\begin{equation}
(C_M)_{a}\rightarrow \frac{\delta }{i\delta (j_M)^{a}}.
\end{equation}%
The total vacuum transition amplitude can  then be written as
\begin{equation}
\langle 0_{+}|0_{-}\rangle ^{A}=\left[ \int \left[ D\Psi \right] \left[ D%
\bar{\Psi}\right] \left[ D(C_M)_{a}\right] \;\exp \left[ i\int d^{4}x\;L_{eff}%
\right] \exp \left[ i\sum_{M,a}\int d^{4}x\;(j_M)^{a}(C_M)_{a}\right] \right]
_{(j_M)^{a}=0}
\end{equation}%
with
\begin{equation}
L_{eff}=\bar{\Psi}\left( i\gamma ^{\mu }\left( \partial _{\mu
}+ieA_{\mu }\right) -m-\sum_{M,a}q_{M}M^{a}\frac{\delta }{i\delta
(j_M)^{a}}\right) \Psi +\sum_{M,a}\left[ \frac{p_{M}}{2}(C_M)_{a}(C_M)^{a}\right].
\end{equation}%
Now we can perform the Gaussian integral over the fields $(C_M)_{a}$,
which includes the linear contribution of the currents, to obtain
\begin{eqnarray}
\langle 0_{+}|0_{-}\rangle ^{A} &=&\int \left[ D\Psi \right] \left[ D\bar{%
\Psi}\right] \;\exp \left[ i\int d^{4}x\;\bar{\Psi}\left( i\gamma
^{\mu }\left( \partial _{\mu }+ieA_{\mu }\right)
-m-\sum_{M,a}q_{M}M^{a}\frac{\delta
}{i\delta (j_M)^{a}}\right) \Psi \right]   \nonumber \\
&&\times \exp \left[ \int
d^{4}x\sum_{a}\frac{i}{p_{M}}\;(j_M)^{a}(j_M)_{a}\right],
\end{eqnarray}%
where the limit $(j_M)_{a}\rightarrow 0\;$is implicit. Notice that in
order for the functional integral of the fields $(C_M)_{a}$  to be well
defined in the usual Euclidean analytic continuation we must require that
$p_{a}>0$.

Using the standard formula (See for example Eq. (1.53) in Ref. \cite{DR})
\begin{equation}
F\left( \frac{\delta }{i\delta j}\right) \exp \left[ \frac{i}{2}\int
j\Delta
j\right] =\left[ \exp \left[ \frac{i}{2}\int j\Delta j\right] \exp \left[ -%
\frac{i}{2}\int \frac{\delta }{\delta X}\Delta \frac{\delta }{\delta X}%
\right] F(X)\right] _{X=\Delta j},
\end{equation}
where $\Delta (x-x^{\prime })$ is  proportional to $\delta (x-x^{\prime
})$  in each case, we can rewrite the above equation as
\begin{eqnarray}
\langle 0_{+}|0_{-}\rangle ^{A} &=&\exp \left[ -i\int d^{4}x\sum_{M,a}\frac{1}{%
2p_{M}}\;\frac{\delta }{\delta (X_M)^{a}}\frac{\delta }{\delta
(X_M)_{a}}\right]
\nonumber \\
&&\times \int d\Psi d\bar{\Psi}\exp \left[ i\int
d^{4}x\;\bar{\Psi}\left( \left( i\gamma ^{\mu }\left( \partial _{\mu
}+ieA_{\mu }\right) -m\right)
-\sum_{M,a}q_{M}M^{a}(X_M)_{a}\right) \Psi \right] _{(X_M)=0}, \nonumber \\
&=&\left[ \exp \left[ -i\int d^{4}x\sum_{M,a}\frac{1}{2p_{M}}\;\frac{\delta }{%
\delta (X_M)^{a}}\frac{\delta }{\delta (X_M)_{a}}\right] \det \left( \left(
i\gamma ^{\mu }\left( \partial _{\mu }+ieA_{\mu }\right) -m\right)
-\sum_{M,a}q_{M}M^{a}(X_M)_{a}\right) \right] _{X_M=0}. \nonumber \\
\end{eqnarray}
where we have already taken the limit $(j_M)^a=0$. The next step is to
rewrite%
\begin{equation}
\left( i\gamma ^{\mu }\left( \partial _{\mu }+ieA_{\mu }\right)
-m\right) -\sum_{M,a}q_{M}M^{a}(X_M)_{a}=\left( i\gamma ^{\mu }\left(
\partial _{\mu }+ieA_{\mu }\right) -m\right) \left(
1-G_{A}\sum_{M,a}q_{M}M^{a}(X_M)_{a}\right) ,
\end{equation}
where  $G_{A}$ is the fermion Green function in the presence of the external
electromagnetic field $A_{\mu }$, satisfying
\begin{equation}
\left( i\gamma ^{\mu }\left( \partial _{\mu }+ieA_{\mu }\right)
-m\right) G_{A}=1.
\end{equation}
Also we recall the expression
\begin{equation}
\det P=\exp \left[ Tr\ln P\right] ,
\end{equation}%
where the trace  is understood both in coordinate as well as
in internal spaces. In this way we have
\begin{eqnarray}
\langle 0_{+}|0_{-}\rangle ^{A} &=&\exp \left[ iW_{EM}^{(1)}\right] \exp
\left[
-i\int d^{4}x\sum_{M,a}\frac{1}{2p_{a}}\;\frac{\delta }{\delta (X_M)^{a}}\frac{%
\delta }{\delta (X_M)_{a}}\right]  \nonumber \\
&&\times \exp \left[ Tr\ln \left( \left(
1-G_{A}\sum_{M,a}q_{M}M^{a}(X_M)_{a}\right) \right) \right] .
\end{eqnarray}%
To make an estimation analogous to the two-loop corrections $%
W_{EM}^{(2)}(A)$ to the effective action $W^{(1)}(A)$, which is
already calculated in Ref. \cite{DR}, we expand the third exponent \
up to terms quadratic in $(X_M)_{a}$.  Calling the operator
\begin{equation}
G_{A}\sum_{M,a}q_{M}M^{a}(X_M)_{a}=\mathcal{G}=\sum_{M}\mathcal{G}_{M}
\label{OPG}
\end{equation}
we are left with
\begin{equation}
\langle 0_{+}|0_{-}\rangle ^{A}=\exp \left[ iW_{EM}^{(1)}\right] \left[ \exp %
\left[ -i\int d^{4}x\sum_{M,a}\frac{1}{2p_{M}}\;\frac{\delta }{\delta (X_M)^{a}}%
\frac{\delta }{\delta (X_M)_{a}}\right] \exp \left[ iTr\left( i\mathcal{G}%
\right) \right] \exp \left[ \frac{i}{2}Tr\left( i\mathcal{G}^{2}\right) %
\right] \right] _{X_M=0}.  \label{VA1}
\end{equation}
With the purpose of having  a preliminary  order of magnitude estimation of the corrections
we next consider separately each of the terms $\mathcal{G}_{M}$ in
the summation appearing in Eq. (\ref{OPG}). In this way we are
neglecting the contribution of the interference terms among different
currents $j_M$, even though their contribution could also be calculated
along the lines presented here.

In order to proceed we make use of another well-known identity
\begin{eqnarray}
\exp \left[ -\frac{i}{2}\int \frac{\delta }{\delta X}A\frac{\delta }{\delta X%
}\right] \exp \left[ \frac{i}{2}\int XBX+i\int FX\right]  &=&\exp \left[ +%
\frac{1}{2}Tr\ln \left( 1-BA\right) ^{-1}\right]   \nonumber \\
&&\times \exp \left[ -\frac{i}{2}\int X\;B(1-AB)^{-1}\;X\right]
\nonumber
\\
&&\times \exp \left[ +i\int X(1-BA)^{-1}F\right]   \nonumber \\
&&\times \exp \left[ +\frac{i}{2}\int FA(1-BA)^{-1}F\right] ,
\end{eqnarray}%
for a fixed $M$. Next we identify the corresponding operators%
\begin{eqnarray}
\left( A_{M}\right)_{\;\;c}^{b}(x-x^{\prime }) &=&\frac{1}{p_{M}}%
\delta _{c}^{b}\delta ^{4}(x-x^{\prime }),\;\
F_{M_{a}}=iq_{M}G_{A}(x,x^{\prime })M_{a}, \\
\left( B_{M}\right) _{\;\;c}^{b}(x,x^{\prime }) &=&i\
q_{M}^{2}G_{A}(x,x^{\prime })M^{b}G_{A}(x^{\prime },x)M_{c}\;,\;\;
\end{eqnarray}%
where $\delta _{c}^{b}$ denotes the identity in the corresponding case.
For example
\begin{equation}
M^a\rightarrow \gamma ^{\mu },\;\delta _{\;c}^{b}\rightarrow
\delta _{\;\nu }^{\mu }\;;\;\;\;M^{a}\rightarrow \sigma ^{\mu \nu
},\;\delta _{\;c}^{b}\rightarrow \frac{1}{2}\left( \delta _{\;\alpha
}^{\mu }\delta _{\;\beta }^{\nu }-\delta _{\;\beta }^{\mu }\delta
_{\;\alpha }^{\nu }\right).
\end{equation}
\begin{table}[tbp]
\begin{tabular}{|l|l|l|l|}
\hline
M &$\,\,\, M^{c}\,\,\,$ & \, \, $\,\,\, \theta_M = \sum_{c} tr(M^{c} M_c) \,\,\, $ & $%
\,\,\, \theta_{3M} = \sum_{c} tr(\sigma_3 M^{c} \sigma_3 M_c) \,\,\,$ \\
\hline
S&\,\,\, $I_{4\times 4}$ & $\hskip1.5cm 4$ & $\hskip1.5cm 4$ \\
V&$\,\,\, \gamma^\mu$ & $\hskip1.5cm 16 $ & $\hskip1.5cm 0$ \\
T&$\,\,\, \sigma^{\mu\nu}$ & $\hskip1.5cm 48 $ & $\hskip1.3cm -16$ \\
PV&$\,\,\, i\gamma^\mu \gamma_5$ & $\hskip1.2cm -16 $ & $\hskip1.6cm 0$ \\
PS&$\,\,\, \gamma_5 $ & $\hskip1.3cm -4 $ & $\hskip1.4cm -4$ \\ \hline
\end{tabular}%
\caption{General notation and traces for the different contributions to $l_{M}$ in
Eq.(\ref{eql}). } \label{table1}
\end{table}
In the sequel we also restrict ourselves to the case of a constant (or
slowly varying) external electromagnetic field. In this case $
G_{A}(x,x^{\prime })$ is independent of position so that the
transformation properties under the Lorenz group imply that
$F_{M_a}=0$. (See for example Eq. (7.8) of Ref. \cite{DR}). This
further simplifies Eq. (\ref{VA1}) to
\begin{equation}
\langle 0_{+}|0_{-}\rangle _{M}^{A}=\exp \left[ iW_{EM}^{(1)}\right] \exp %
\left[ +\frac{1}{2}Tr\ln \left( 1-B_{M}A_{M}\right) ^{-1}\right] \simeq \exp %
\left[ iW_{EM}^{(1)}\right] \exp \left[ +\frac{1}{2}Tr\left(
B_{M}A_{M}\right) \right]
\end{equation}%
This defines the sought correction
\begin{equation}
\exp \left[ iW_{M}^{(2)}\right] =\exp \left[
+\frac{1}{2}Tr\left( B_{M}A_{M}\right) \right]
,\;\;\;M\;fixed
\end{equation}%
where
\begin{equation}
Tr\left( B_{M}A_{M}\right) =i\ \frac{r_{M}}{\Lambda
^{2}}\sum_c \int d^{4}x\;tr\left( G_{A}(x,x)M^{c}G_{A}(x,x)M_{c}\right)
,\;\;\;\;
\end{equation}%
where $tr$ is the trace in the Dirac-matrices space. In other words
we have
\begin{equation}
W_{M}^{(2)}=\frac{2r_{M}}{\Lambda ^{2}}\sum_c \int  d^{4}x\;tr\left(
G_{A}(x,x)M^{c}G_{A}(x,x)M_{c}\right) \equiv \frac{r_{M}}{\Lambda
^{2}}\int d^{4}x\;l_{M}.  \label{W2CURR}
\end{equation}%
In order to estimate each contribution we further consider the case
where the external field is a constant magnetic $B$ field in the
$z$-direction. We avoid the constant electric field case because of  the
inherent instability due to pair creation  in the strong field
regime. For the case under consideration we have \cite{JS}
\begin{equation}
G_{A}(x,x)=\frac{m}{16\pi ^{2}}\int_{0}^{\infty }\frac{ds}{s^{2}}\;\exp %
\left[ -ism^{2}\right] \frac{z}{\sin z}e^{i\sigma
_{3}z},\;\;\;s=eBs,\;\sigma _{3}=i\gamma ^{1}\gamma ^{2}\;\;,
\end{equation}
where $m$ is the mass of the integrated fermion. In this way, the
general expression for the quantity $l_{M}$ defined in Eq.
(\ref{W2CURR}) is
\begin{eqnarray}
l_{M} &=&\left( \frac{m}{16\pi ^{2}}\right) ^{2}\int_{0}^{\infty }\frac{ds}{%
s^{2}}\;\frac{ds^{\prime }}{s^{\prime 2}}\exp \left[ -ism^{2}\right] \exp %
\left[ -is^{\prime }m^{2}\right] \frac{z}{\sin z}\frac{z^{\prime
}}{\sin
z^{\prime }}  \nonumber \\
&&\times \sum_c \left[ \left( \cos z\cos z^{\prime }tr\left(
M^{c}M_{c}\right) -\sin z\sin z^{\prime }tr\left( \sigma
_{3}M^{c}\sigma _{3}M_{c}\right) \right) +\sin (z-z^{\prime
})tr(\sigma _{3}M^{c}M_{c})\right]   \label{eql}
\end{eqnarray}%
The last term in the second line of Eq.(\ref{eql}) vanishes because
$\sum_c  M^{c}M_{c}$ is always a multiple of the identity
and $tr(\sigma _{3})=0.\;$Let us introduce the notation
\begin{equation}
\theta _{M}=tr\left(\sum_c M^{c}M_{c}\right) ,\;\;\;\theta _{3M}=tr\left(\sum_c
\sigma _{3}M^{c}\sigma _{3}M_{c}\right) ,\;\;\;\;
\end{equation}%
which values are given in Table \ref{table1}. Also it is convenient to define
\begin{equation}
I_{1}=\int_{0}^{\infty }\frac{ds}{s^{2}}\;\exp \left[ -ism^{2}\right] \frac{%
z\cos z}{\sin z},\;\;\;\;I_{2}=\int_{0}^{\infty }\frac{ds}{s^{2}}\;\exp %
\left[ -ism^{2}\right] z=eB\int_{0}^{\infty }\frac{ds}{s}\;\exp
\left[ -ism^{2}\right].
\end{equation}%
In terms of the above quantities we have
\begin{equation}
l_{M}=\left( \frac{m}{16\pi ^{2}}\right) ^{2}\left[ \theta
_{M}I_{1}^{2}\;-\theta _{3M}I_{2}^{2}\right].
\end{equation}%
The second contribution, with infinite coefficient, behaves as $B^{2}$ so that we consider it
as part of the renormalization procedure that we demand of the low
energy limit of the
effective potential
\begin{equation}
\lim_{B\rightarrow 0}\;V_{eff}(B)=-\rho B^{2}, \qquad \rho > 0.
\end{equation}%
In this way we consider the full expression for the contribution due to
the additional currents to be
\begin{eqnarray}
L^{(2)} =\sum_{M}L_{M}^{(2)}, \qquad L_{M}^{(2)} =\left( \frac{m}{16\pi ^{2}}\right)
^{2}\frac{r_{M}}{\Lambda
^{2}}\;\theta _{M}\;I_{1}^{2}.  \label{FINALCORR}
\end{eqnarray}
The effective Lagrangians in  (\ref%
{FINALCORR})  lead to the effective potentials%
\begin{equation}
V_{M}^{(2)}=-L_{M}^{(2)}=-\left( \frac{m}{16\pi ^{2}}%
\right) ^{2}\;\frac{r_{M}}{\Lambda ^{2}}\, \theta _{M}\;I_{1}^{2}.
\label{FINALVEFF}
\end{equation}%
The main point to be made is that each contribution to the effective potential arising from the subspace $M$  contributes with an specific sign, as can be seen
from Table 1. In these way, as it will be argued in the sequel,
it is possible to have a net positive contribution for the total
effective potential in the limit $B/B_{0}\rightarrow \infty$,
where $B_{0}=m^{2}/e$ is the critical value of the magnetic field.
Let us emphasize that $V_M^{(2)}$ is still not well defined because
the integral $I_{1}$ diverges. This means that we should regularize
it and further implement a renormalization prescription
for the full effective Lagrangian%
\begin{equation}
L_{eff}=\alpha B^{2}+L_{EM}^{(1)}(B)+L_{EM}^{(2)}(B)+L^{(2)}(B).
\end{equation}%
Here $L_{EM}^{(2)}(B)$ refers to the two-loop standard electromagnetic correction.
Our renormalization conditions will be%
\begin{equation}
\lim_{B\rightarrow 0}L_{eff}=\rho B^2, \qquad \rho > 0,
\end{equation}%
in such a way that the effective potential is a decreasing function
in the
vicinity of $B=0.$ The final normalization to the value of $%
L_{eff}=-b^{2}/2, b \rightarrow 0 $, will be imposed at the end of the procedure,
after we expand the magnetic field around the expected stable minimum $B_{Min}$ of
$V_{eff}$, such that
\begin{equation}
B=B_{Min}+b
\end{equation}%
and we are left with the physical component $b$. The renormalization
procedure should be analogous to the one carried out in Ref. \cite{DR}
for the contribution $L_{EM}^{(2)}(A)$ and it is rather involved due
to the complicated field dependence of the regularized version of
the divergent integrals. This calculation is out of the scope of these preliminary
estimations.

Thus we will only consider a particular regularization of
$I_{1}$ to proceed with the comparison of the different high-intensity field
contributions to the effective potential. We take%
\begin{equation}
I_{1}\rightarrow I_{1}=\int_{0}^{\infty }\frac{ds}{s^{2}}\;\exp
\left[ -ism^{2}\right] \left[ \frac{z\cos z}{\sin z}-1\right],
\label{I1REG}
\end{equation}%
which produces a zero contribution when $B\rightarrow 0$. This
integral can be readily calculated from the first order
electromagnetic correction of the effective action given by \cite{DR}
\begin{equation}
L_{EM}^{(1)}=\frac{1}{8\pi ^{2}}\int_{0}^{\infty }\frac{ds}{s^{3}}\exp
\left[ -ism^{2}\right] \left[ \frac{z\cos z}{\sin
z}+\frac{1}{3}z-1\right] .
\end{equation}%
The result is
\begin{equation}
I_{1}=i\left[ 8\pi ^{2}\frac{\partial L_{EM}^{(1)}}{\partial m^{2}}+\frac{eB^{2}}{%
3m^{2}}\right] ,
\end{equation}%
which yields our final result
\begin{equation}
V_{M}^{(2)}=\left( \frac{m}{16\pi ^{2}}\right) ^{2}\;%
\frac{r_{M}}{\Lambda ^{2}} \,\, \theta _{M}\;\left[ 8\pi ^{2}\frac{\partial L_{EM}^{(1)}%
}{\partial m^{2}}+\frac{eB^{2}}{3m^{2}}\right] ^{2}.
\label{VEFFFIN}
\end{equation}%
In order to produce the required results in the low (high)-intensity field approximations, we list here the behavior of the relevant
terms  obtained from Ref.
\cite{DR}. For $B\rightarrow \infty $ we have
\begin{eqnarray}
V_{EM}^{(1)} &\rightarrow &-\frac{e^{2}}{24\pi ^{2}}B^{2}\ln \left( \frac{eB}{%
m^{2}}\right),   \label{V1EMINF} \\
V_{EM}^{(2)} &\rightarrow &-\frac{e^{4}}{128\pi ^{4}}B^{2}\ln \left( \frac{eB%
}{m^{2}}\right),   \label{V2EMINF} \\
8\pi ^{2}\frac{\partial L_{EM}^{(1)}}{\partial
m^{2}}+\frac{eB^{2}}{3m^{2}} &\rightarrow &eB\ln \left(
\frac{2eB}{m^{2}}\right).   \label{V2JINF}
\end{eqnarray}%
The limit $B\rightarrow 0$ produces
\begin{eqnarray}
V_{EM}^{(1)} &\rightarrow &-\frac{e^{4}}{360\pi ^{2}m^{4}}B^{4},
\label{V1EMCERO}
\\
8\pi ^{2}\frac{\partial L_{EM}^{(1)}}{\partial
m^{2}}+\frac{eB^{2}}{3m^{2}}
&\rightarrow &\left( \frac{eB^{2}}{3m^{2}}+\frac{1}{18}\frac{e^{2}B^{4}}{%
m^{4}}\right).   \label{VJCERO}
\end{eqnarray}%
In this way, the final contributions of a given current $%
j_M$ to the effective potential are
\begin{eqnarray}
B &\rightarrow &\infty : \qquad V_{M}^{(2)}=\left( \frac{1%
}{16\pi ^{2}}\right) ^{2}\left( \frac{m}{\Lambda }\right)
^{2}r_{M}\, \theta
_{M}\;e^{2}B^{2}\left[ \ln \left( \frac{2eB}{m^{2}}\right) \right] ^{2}, \\
B &\rightarrow &0: \qquad V_{M}^{(2)}=\left( \frac{1}{%
16\pi ^{2}}\right) ^{2}\left( \frac{m}{\Lambda }\right)
^{2}r_{M}\, \theta _{M}\left(
\frac{eB^{2}}{3m^{2}}+\frac{1}{18}\frac{e^{2}B^{4}}{m^{4}}\right)^2.
\end{eqnarray}
The total contribution from the currents $j_M$ is
\begin{equation}
V^{(2)}=\sum_M V_{M}^{(2)}.
\label{TOTALCONTR}
\end{equation}

\section*{Acknowledgments}

The work of J.A. was partially supported by Fondecyt \# 1060646. LFU
acknowledges support from the grants DAGPA-UNAM-IN 109107 and CONACyT \#
55310. LFU also acknowledges R. Lehnert for useful suggestions and comments.

\end{document}